\documentclass[prl,reprint,twocolumn,letterpaper,groupedaddress,amsmath,amssymb,showpacs,aps,english,superscriptaddress]{revtex4-1}

\usepackage{subfigure}
\usepackage{graphicx}
\usepackage{bm,mathrsfs}
\usepackage{amsmath}
\usepackage{amssymb}
\usepackage{esint}
\usepackage{setspace}
\usepackage{hyperref}
\usepackage{color,colortbl}
\usepackage{colortbl}
\usepackage{mathtools}
\usepackage{multirow}
\usepackage{bbold}
\usepackage{pbox}
\usepackage{todonotes}
\usepackage{hhline}
\usepackage{comment}
\usepackage{tabularx}
\usepackage{cancel}
\usepackage{placeins}
\allowdisplaybreaks
\newcolumntype{C}[1]{>{\centering\arraybackslash}p{#1}}

\makeatletter

\begin{document}

\title{Silicon vacancy center in 4H-SiC: Electronic structure and spin-photon interfaces}
\author{\"O. O. Soykal}
\affiliation{NRC post doc residing at Code 6877 Naval Research Laboratory, Washington, D.C. 20375, USA}
\author{Pratibha Dev}
\affiliation{NRC post doc residing at Code 6877 Naval Research Laboratory, Washington, D.C. 20375, USA}
\affiliation{Department of Physics and Astronomy, Howard
University, Washington, DC, USA}
\author{Sophia E. Economou}
\affiliation{Naval Research Laboratory, Washington, DC 20375}
\affiliation{Department of Physics, Virginia Tech, Blacksburg, Virginia 24061, USA}
\begin{abstract}
Defects in silicon carbide are of intense and increasing interest for quantum-based applications due to this material's properties and technological maturity. We calculate the multi-particle symmetry adapted wave functions of the negatively charged silicon vacancy defect in hexagonal silicon carbide via use of group theory and density functional theory and find the effects of spin-orbit and spin-spin interactions on these states. Although we focused on $\textrm{V}_{\textrm{Si}}^-$ in 4H-SiC, because of its unique fine structure due to odd number of active electrons, our methods can be easily applied to other defect centers of different polytpes, especially to the 6H-SiC. Based on these results we identify the mechanism that polarizes the spin under optical drive, obtain the ordering of its dark doublet states, point out a path for electric field or strain sensing, and find the theoretical value of its ground-state zero field splitting to be 68 MHz, in good agreement with experiment. Moreover, we present two distinct protocols of a spin-photon interface based on this defect. Our results pave the way toward novel quantum information and quantum metrology applications with silicon carbide.
\end{abstract}

\maketitle

Over the last several years, deep-center defects in solids have been intensely researched for applications in quantum information \cite{Togan_nature11,Bernien_Nature13}, quantum sensing and nanoscale imaging \cite{Grinolds_NP13} including bioimaging \cite{Balasubramanian_Nature08,Shi_science15}. Their success and popularity stem from their unique properties, combining advantages from atomic and solid state systems—-most notably long spin coherence times even at room temperature and integrability into a solid state matrix. The NV center in diamond is the most studied defect for quantum technologies, so that its properties, strengths and limitations are by now very well understood. Deep defect centers in silicon carbide (SiC) have emerged as strong contenders due to this material's significantly lower cost, availability of mature microfabrication technologies \cite{Song_OptExp11,Maboudian_JVSci13}, and favorable optical emission wavelengths \cite{Baranov_prb11}.

Some of the stable defects in SiC have the same structure as the NV center in diamond in terms of symmetry and the number of active electrons and, as a result, spin and electronic structure. Such defects include the silicon-carbon divacancy, which has been investigated over the last several years \cite{Son_PRL06,Koehl_nature11,Falk_natcom13,Christle_nature14}. Experiments \cite{Mizuochi2002,Baranov_prb11,Riedel_PRL12,Kraus_NP14,Kraus_scirep14,Widmann_nmat15,Carter_PRBRC15} on the Si \emph{monovacancy} ($\textrm{V}_{\textrm{Si}}^-$) have shown that this is a distinct defect in terms of electronic and spin structure. It features a ground state with total spin $3/2$ \cite{Mizuochi2002,Kraus_NP14}, offering both quantitative improvements and qualitatively new capabilities \cite{Kraus_scirep14} compared to NV-like defects. To date, room temperature spin polarization and coherent control of $\textrm{V}_{\textrm{Si}}^-$ have been implemented via electron spin resonance \cite{Soltamov_PRL12,Widmann_nmat15} and optically detected magnetic resonance (ODMR) \cite{Sorman_PRB2000,Baranov_prb11,Kraus_NP14,Carter_PRBRC15}. Unlike the well-studied NV center in diamond \cite{Lenef_PRB96,maze_njp11,doherty_njp11}, theoretical studies of the $\textrm{V}_{\textrm{Si}}^-$ in SiC have been mostly limited to finding single-particle levels and their energies via density functional theory (DFT) \cite{janzen_physicab09,Weber_PNAS10,gali_JMatR12}. While such DFT calculations are an important first step, it is of crucial importance to obtain the multi-particle electronic structure to understand the properties of this defect and take full advantage of the novel opportunities it affords.

\begin{figure}[!hb]
  \centering
\includegraphics[width=8.0cm]{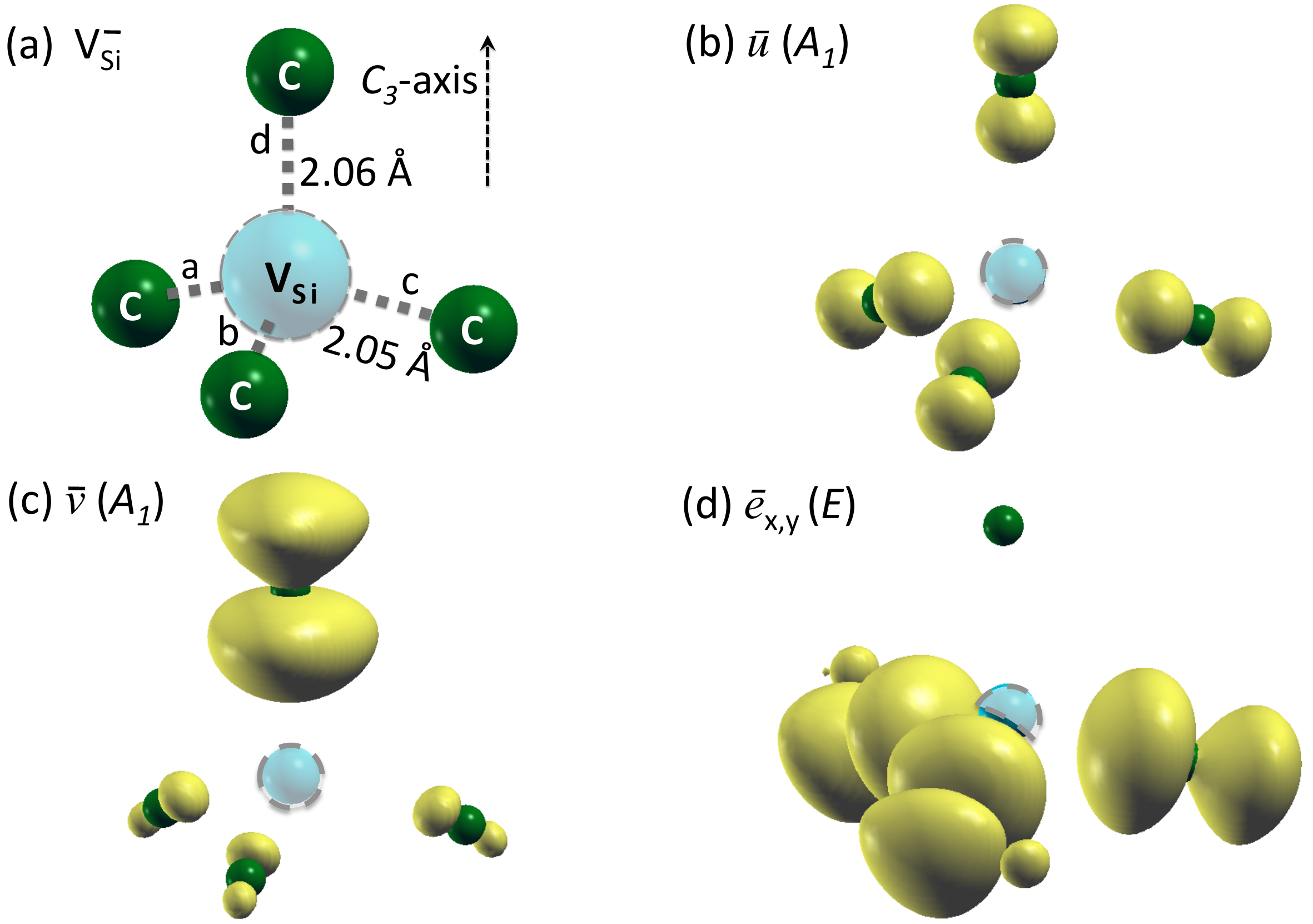}		
  \caption{(color online) $\textrm{V}_{\textrm{Si}}^-$ in 4H-SiC: (a) $C_{3\nu}$-structure of the defect, and the optically-active orbitals of $\textrm{V}_{\textrm{Si}}^-$ using DFT: (b) $\bar{u}$ ($A_1$ symmetry), (c) $\bar{v}$ ($A_1$), and (d) $\bar{e}_{x,y}$ ($E$). Only carbons near the $\textrm{V}_{\textrm{Si}}^-$ are shown for clarity.}
  \label{Fig1}
\end{figure}

\setlength{\tabcolsep}{5pt}
\renewcommand{\arraystretch}{1.2}
\begin{table*}[ht]
\centering
\begin{tabular}{|c||c|c|c|c|l|c|}
\hline
Orbital & $S$ & $m_s$ & $\Gamma$ & $\Gamma_o\otimes\Gamma_s$ & \multicolumn{1}{c|}{Symmetry adapted total wave functions} & $\textrm{Label}$ \\
\hline\hline
 &  &  & \multirow{-1}{*}{$\prescript{1}{}E_{3/2}$} & \multirow{-1}{*}{$A_2\otimes\prescript{2}{}E_{3/2}$} & \multirow{-1}{*}{$\begin{array}{l}||ve_xe_y+i\bar{v}\bar{e}_x\bar{e}_y\rangle/\sqrt{2}\end{array}$} & $\Psi_\textrm{g}^1$ \\ \cline{4-7}

 &  & \multirow{-2}{*}{$\pm\frac{3}{2}$} & \multirow{-1}{*}{$\prescript{2}{}E_{3/2}$} & \multirow{-1}{*}{$A_2\otimes\prescript{1}{}E_{3/2}$} & \multirow{-1}{*}{$\begin{array}{l}||ve_xe_y-i\bar{v}\bar{e}_x\bar{e}_y\rangle/\sqrt{2}\end{array}$} & $\Psi_\textrm{g}^2$ \\ \cline{3-7}

 &  & \multirow{-1}{*}{$+\frac{1}{2}$} & \multirow{-1}{*}{$E_{1/2}^+$} &  & $\begin{array}{l}||ve_x\bar{e}_y+v\bar{e}_xe_y+\bar{v}e_xe_y\rangle/\sqrt{3}\end{array}$ & $\Psi_\textrm{g}^3$ \\ \cline{3-4}\cline{6-7}

\multirow{-4}{*}{\pbox{5cm}{$ve^2$ \\ (\textrm{Ground}) \\ \& \\ $ue^2$ (\textrm{q1})}} & \multirow{-4}{*}{$\frac{3}{2}$} &  \multirow{-1}{*}{$-\frac{1}{2}$} & \multirow{-1}{*}{$E_{1/2}^-$} & \multirow{-2}{*}{$A_2\otimes E_{1/2}$} & $\begin{array}{l}||\bar{v}\bar{e}_xe_y+\bar{v}e_x\bar{e}_y+v\bar{e}_x\bar{e}_y\rangle/\sqrt{3}\end{array}$ & $\Psi_\textrm{g}^4$ \\ 

\hline\hline

& & \multirow{-1}{*}{$+\frac{1}{2}$} & $E_{1/2}^+$ & & \multirow{-1}{*}{$\begin{array}{l}||e_x\bar{e}_xe_y+i e_y\bar{e}_ye_x\rangle/\sqrt{2}\end{array}$} & $\Psi_{\textrm{d}1}^1$ \\ \cline{3-4}\cline{6-7}

& & \multirow{-1}{*}{$-\frac{1}{2}$} & $E_{1/2}^-$ & & \multirow{-1}{*}{$\begin{array}{l}||\bar{e}_xe_x\bar{e}_y-i\bar{e}_ye_y\bar{e}_x\rangle/\sqrt{2}\end{array}$} & $\Psi_{\textrm{d}1}^2$ \\ \cline{3-4}\cline{6-7}

& & & \multirow{-1}{*}{$\prescript{1}{}E_{3/2}$} & & \multirow{-1}{*}{$\begin{array}{l}||(e_x\bar{e}_xe_y-i e_y\bar{e}_ye_x)-i(\bar{e}_xe_x\bar{e}_y+i\bar{e}_ye_y\bar{e}_x)\rangle/2\end{array}$} & \multirow{-1}{*}{$\Psi_{\textrm{d}1}^3$} \\ \cline{4-4}\cline{6-7}

\multirow{-4}{*}{\pbox{5cm}{$e^3$ \\ (\textrm{d1})}} & \multirow{-4}{*}{$\frac{1}{2}$} & \multirow{-2}{*}{$\pm\frac{1}{2}$} & \multirow{-1}{*}{$\prescript{2}{}E_{3/2}$} & \multirow{-4}{*}{$E\otimes E_{1/2}$} & \multirow{-1}{*}{$\begin{array}{l}||(e_x\bar{e}_xe_y-i e_y\bar{e}_ye_x)+i(\bar{e}_xe_x\bar{e}_y+i\bar{e}_ye_y\bar{e}_x)\rangle/2\end{array}$} & \multirow{-1}{*}{$\Psi_{\textrm{d}1}^4$} \\

\hline\hline

 &  & \multirow{-1}{*}{$+\frac{1}{2}$} & $E_{1/2}^+$ &  & $\begin{array}{l}||ve_x\bar{e}_y+v\bar{e}_xe_y-2\bar{v}e_xe_y\rangle/\sqrt{6}\end{array}$ & $\Psi_{\textrm{d}2}^1$ \\ \cline{3-4}\cline{6-7}

\multirow{-2}{*}{\pbox{5cm}{$ve^2$ \\ (\textrm{d2})}} & \multirow{-2}{*}{$\frac{1}{2}$} & \multirow{-1}{*}{$-\frac{1}{2}$} & $E_{1/2}^-$ & \multirow{-2}{*}{$A_2\otimes E_{1/2}$} & $\begin{array}{l}||\bar{v}\bar{e}_xe_y+\bar{v}e_x\bar{e}_y-2v\bar{e}_x\bar{e}_y\rangle/\sqrt{6}\end{array}$ & $\Psi_{\textrm{d}2}^2$ \\ 

\hline\hline

 & & & & & & \\
 & & & \multirow{-2}{*}{$\prescript{1}{}E_{3/2}$} & & \multirow{-2}{*}{\renewcommand\arraystretch{0.1}$\begin{array}{l}||(ve_x\bar{e}_y-v\bar{e}_xe_y)-i(\bar{v}\bar{e}_xe_y-\bar{v}e_x\bar{e}_y) \\ +i(ve_x\bar{e}_x-ve_y\bar{e}_y)-(\bar{v}\bar{e}_xe_x-\bar{v}\bar{e}_ye_y)\rangle/2\sqrt{2}\end{array}$} & \multirow{-2}{*}{$\Psi_{\textrm{d}3}^1$}\\ \cline{4-4}\cline{6-7}
 & & & & & & \\
 &  &  & \multirow{-2}{*}{$\prescript{2}{}E_{3/2}$} &  & \multirow{-2}{*}{\renewcommand\arraystretch{0.1}$\begin{array}{l}||(ve_x\bar{e}_y-v\bar{e}_xe_y)+i(\bar{v}\bar{e}_xe_y-\bar{v}e_x\bar{e}_y) \\ +i(ve_x\bar{e}_x-ve_y\bar{e}_y)+(\bar{v}\bar{e}_xe_x-\bar{v}\bar{e}_ye_y)\rangle/2\sqrt{2}\end{array}$} & \multirow{-2}{*}{$\Psi_{\textrm{d}3}^2$} \\ \cline{4-4}\cline{6-7}
 &  &  & \multirow{-1}{*}{$E_{1/2}^+$} &  & \multirow{-1}{*}{\renewcommand\arraystretch{0.1}$\begin{array}{l}||(ve_x\bar{e}_y-v\bar{e}_xe_y)-i(ve_x\bar{e}_x-ve_y\bar{e}_y)\rangle/2\end{array}$} & \multirow{-1}{*}{$\Psi_{\textrm{d}3}^3$} \\ \cline{4-4}\cline{6-7}
\multirow{-6}{*}{\pbox{5cm}{$ve^2$ \\ (\textrm{d3}) }} & \multirow{-6}{*}{$\frac{1}{2}$} & \multirow{-6}{*}{$\pm\frac{1}{2}$} & \multirow{-1}{*}{$E_{1/2}^-$} & \multirow{-6}{*}{$E\otimes E_{1/2}$} & \multirow{-1}{*}{\renewcommand\arraystretch{0.1}$\begin{array}{l}||(\bar{v}\bar{e}_xe_y-\bar{v}e_x\bar{e}_y)+i(\bar{v}\bar{e}_xe_x-\bar{v}\bar{e}_ye_y)\rangle/2\end{array}$} & \multirow{-1}{*}{$\Psi_{\textrm{d}3}^4$} \\

\hline\hline

 &  & \multirow{-1}{*}{$+\frac{1}{2}$} & $E_{1/2}^+$ &  & $\begin{array}{l}||ve_x\bar{e}_x+ve_y\bar{e}_y\rangle/\sqrt{2}\end{array}$ & $\Psi_{\textrm{d}4}^1$ \\ \cline{3-4}\cline{6-7}

\multirow{-2}{*}{\pbox{5cm}{$ve^2$ \\ (\textrm{d4})}} & \multirow{-2}{*}{$\frac{1}{2}$} & \multirow{-1}{*}{$-\frac{1}{2}$} & $E_{1/2}^-$ & \multirow{-2}{*}{$A_1\otimes E_{1/2}$} & $\begin{array}{l}||\bar{v}\bar{e}_xe_x+\bar{v}\bar{e}_ye_y\rangle/\sqrt{2}\end{array}$ & $\Psi_{\textrm{d}4}^2$ \\ 

\hline\hline

& & \multirow{-1}{*}{$+\frac{1}{2}$} & $E_{1/2}^+$ & & \multirow{-1}{*}{$\begin{array}{l}||v\bar{v}e_x-i v\bar{v}e_y\rangle/\sqrt{2}\end{array}$} & $\Psi_{\textrm{d}5}^1$ \\ \cline{3-4}\cline{6-7}

& & \multirow{-1}{*}{$-\frac{1}{2}$} & $E_{1/2}^-$ & & \multirow{-1}{*}{$\begin{array}{l}||\bar{v}v\bar{e}_x+i\bar{v}v\bar{e}_y\rangle/\sqrt{2}\end{array}$} & $\Psi_{\textrm{d}5}^2$ \\ \cline{3-4}\cline{6-7}

& & & \multirow{-1}{*}{$\prescript{1}{}E_{3/2}$} & & \multirow{-1}{*}{$\begin{array}{l}||(v\bar{v}e_x+i v\bar{v}e_y)+i(\bar{v}v\bar{e}_x-i\bar{v}v\bar{e}_y)\rangle/2\end{array}$} & \multirow{-1}{*}{$\Psi_{\textrm{d}5}^3$} \\ \cline{4-4}\cline{6-7}

\multirow{-4}{*}{\pbox{5cm}{$v^2e$ \\ (\textrm{d5})}} & \multirow{-4}{*}{$\frac{1}{2}$} & \multirow{-2}{*}{$\pm\frac{1}{2}$} & \multirow{-1}{*}{$\prescript{2}{}E_{3/2}$} & \multirow{-4}{*}{$E\otimes E_{1/2}$} & \multirow{-1}{*}{$\begin{array}{l}||(v\bar{v}e_x+iv\bar{v}e_y)-i(\bar{v}v\bar{e}_x-i\bar{v}v\bar{e}_y)\rangle/2\end{array}$} & \multirow{-1}{*}{$\Psi_{\textrm{d}5}^4$} \\

\hline\hline

& & \multirow{-1}{*}{$+\frac{3}{2}$} & $E_{1/2}$ & $E\otimes \prescript{1}{}E_{3/2}$ & $\begin{array}{l}||uve_x\rangle\end{array}$, $\begin{array}{l}||uve_y\rangle\end{array}$ & $\Psi_{\textrm{q}2}^1$, $\Psi_{\textrm{q}2}^2$ \\ \cline{3-7}

& & \multirow{-1}{*}{$-\frac{3}{2}$} & $E_{1/2}$ & $E\otimes \prescript{2}{}E_{3/2}$ & $\begin{array}{l}||\bar{u}\bar{v}\bar{e}_x\rangle\end{array}$, $\begin{array}{l}||\bar{u}\bar{v}\bar{e}_y\rangle\end{array}$ & $\Psi_{\textrm{q}2}^3$, $\Psi_{\textrm{q}2}^4$ \\ \cline{3-7}

& & & \multirow{-1}{*}{$E_{1/2}^+$} & & \multirow{-1}{*}{\renewcommand\arraystretch{0.1}$\begin{array}{l}||(uv\bar{e}_y+u\bar{v}e_y+\bar{u}ve_y)+i(uv\bar{e}_x+u\bar{v}e_x+\bar{u}ve_x)\rangle/\sqrt{6}\end{array}$} & \multirow{-1}{*}{$\Psi_{\textrm{q}2}^5$} \\ \cline{4-4}\cline{6-7}

& & & \multirow{-1}{*}{$E_{1/2}^-$} & & \multirow{-1}{*}{\renewcommand\arraystretch{0.1}$\begin{array}{l}||(\bar{u}\bar{v}e_y+\bar{u}v\bar{e}_y+u\bar{v}\bar{e}_y)-i(\bar{u}\bar{v}e_x+\bar{u}v\bar{e}_x+u\bar{v}\bar{e}_x)\rangle/\sqrt{6}\end{array}$} & \multirow{-1}{*}{$\Psi_{\textrm{q}2}^6$} \\ \cline{4-4}\cline{6-7}		
& & & & & & \\
& & & \multirow{-2}{*}{$\prescript{1}{}E_{3/2}$} & & \multirow{-2}{*}{\renewcommand\arraystretch{0.1}$\begin{array}{l}||(uv\bar{e}_y+u\bar{v}e_y+\bar{u}ve_y)-i(u\bar{v}\bar{e}_y+\bar{u}v\bar{e}_y+\bar{u}\bar{v}e_y) \\ -i(uv\bar{e}_x+u\bar{v}e_x+\bar{u}ve_x)+(u\bar{v}\bar{e}_x+\bar{u}v\bar{e}_x+\bar{u}\bar{v}e_x)\rangle/2\sqrt{3}\end{array}$} & \multirow{-2}{*}{$\Psi_{\textrm{q}2}^7$} \\ \cline{4-4}\cline{6-7}
& & & & & & \\
\multirow{-8}{*}{\pbox{5cm}{$uve$ \\ (\textrm{q2})}} & \multirow{-8}{*}{$\frac{3}{2}$} & \multirow{-6}{*}{$\pm\frac{1}{2}$} & \multirow{-2}{*}{$\prescript{2}{}E_{3/2}$} & \multirow{-6}{*}{$E\otimes E_{1/2}$} & \multirow{-2}{*}{\renewcommand\arraystretch{1}$\begin{array}{l}||(uv\bar{e}_y+u\bar{v}e_y+\bar{u}ve_y)+i(u\bar{v}\bar{e}_y+\bar{u}v\bar{e}_y+\bar{u}\bar{v}e_y) \\ -i(uv\bar{e}_x+u\bar{v}e_x+\bar{u}ve_x)-(u\bar{v}\bar{e}_x+\bar{u}v\bar{e}_x+\bar{u}\bar{v}e_x)\rangle/2\sqrt{3}\end{array}$} & \multirow{-2}{*}{$\Psi_{\textrm{q}2}^8$} \\

\hline

\end{tabular}

\caption{Negatively charged Si vacancy wave functions for various configurations in the three hole representation. The states are classified in terms of orbital electronic configuration, total spin ($S$) and spin projection along the $C_3$-axis ($m_s$), overall symmetry representation of the state ($\Gamma$) and its decomposition in terms of the orbital and spin symmetries ($\Gamma_o\otimes\Gamma_s$). q1 states (not explicitly shown) are defined similarly to states $\Psi_\textrm{g}^1$-$\Psi_\textrm{g}^4$ with the replacement $v\rightarrow u$. The notation $||\dots\rangle$ represents the Slater determinant of each component inside the bracket. The bar (no bar) over an orbital indicates spin down (up).}
\label{Table3}
\end{table*}

In this Letter we address this need by calculating the multi-particle wave functions of $\textrm{V}_{\textrm{Si}}^-$ through a combination of group theory and DFT. We explicitly find the ground states as well as the excited state manifolds, considering both the orbital and the spin degrees of freedom. Furthermore, we investigate the effects of spin-orbit and spin-spin interactions. Based on these results we (i) explain quantitatively the spin polarization mechanism in experiments, (ii) find the zero-field splitting, in good agreement with experiment, (iii) present a mechanism that allows this defect to be used for electric field or strain sensing, and (iv) propose two spin-photon interface protocols enabled by the rich electronic structure of this defect, including the generation of strings of entangled photons and the creation of a Lambda system with potential applications in quantum technologies.

The $C_{6\nu}$ symmetry of bulk 4H-SiC is lowered to the $C_{3\nu}$ point group in the presence of $\textrm{V}_{\textrm{Si}}^-$. The local geometry of $\textrm{V}_{\textrm{Si}}^-$ is shown in Fig.\,\ref{Fig1}(a), where the missing silicon leaves four dangling bonds ($sp^3$-orbitals) on the surrounding carbons. Single electron molecular orbitals (MO) can be constructed from symmetry-adapted linear combinations of the three equivalent $sp^3$-orbitals ($a$, $b$ and $c$) from the basal-plane carbons and the $sp^3$-orbital, $d$, belonging to the carbon atom on the $C_3$-axis that coincides with the crystalline $c$-axis. Using the standard projection operator technique \cite{Tinkham2003} and our DFT results as a guide [Fig.\,\ref{Fig1}(b)-(d)], we obtain the following MOs of the defect center: $u{=}\alpha_u (a+b+c){+}\beta_u d$, $v{=}\alpha_v (a+b+c){+}\beta_v d$, $e_x{=}\alpha_x (2c-a-b)$, and $e_y{=}\alpha_y (a-b)$, 
where the coefficients 
are given in \cite{supplement}. The orbitals, as calculated by DFT, are shown in Fig.\ref{Fig1}. The functions $u$ and $v$ transform as $A_1$, $e_X$ and $e_Y$ transform as the $x$ and $y$ components of the $E$ representation respectively and the states are listed in order of increasing energy according to our DFT calculations.

The electronic configuration of this defect is modeled by three holes, a simpler but equivalent picture to that of five active electrons. Then, the three-hole lowest energy quartet configurations are identified as $ve_xe_y$, $ue_xe_y$, and $uve_x$ (or $uve_y$), respectively, increasing in energy \cite{supplement}. The tensor products of $u$, $v$, and $e_{x,y}$ states with the total spin eigenstates comprise our basis set, from which we calculate the multi-particle symmetry-adapted states compatible with $C_{3\nu}$. The odd number of particles here results in a much more complicated structure compared to NV centers in diamond and divacancies in SiC. Thus, we obtain the multi-particle wave functions systematically by use of the projection operator on the basis states for both the orbital and the spin degree of freedom:
\begin{equation}
\mathcal{P}^{(j)}=(I_j/h)\sum_R\chi^{(j)}(R)^*\Gamma^{(j)}(R),\label{2}
\end{equation}
where, $\chi^{(j)}(R)$ is the character of operation $R$ in the $j^{\text{th}}$ irreducible representation \cite{supplement}, and $\Gamma$ is the irreducible matrix representation for the $R$ symmetry operator (tensor product of the three-particle orbital and spin operators \cite{supplement}). The resulting symmetry adapted states are shown in Table\,\ref{Table3}, and are characterized by the total spin $S$, the orbital and spin symmetry, as well as their overall symmetry. These classifications are of key importance in understanding the nature of these states, their additional interactions, as well as the allowed optical or spin-orbit assisted transitions and selection rules. The ground state manifold has $S{=}3/2$ (quartet), while there are nearby additional manifolds (each a doublet, $S{=}1/2$) with some having the same orbital composition as the ground state and split from each other only due to Coulomb interactions (see Fig.\,\ref{Fig3} and \cite{supplement}).

\begin{figure}[ht]
  \centering
  \includegraphics[width=8.5 cm]{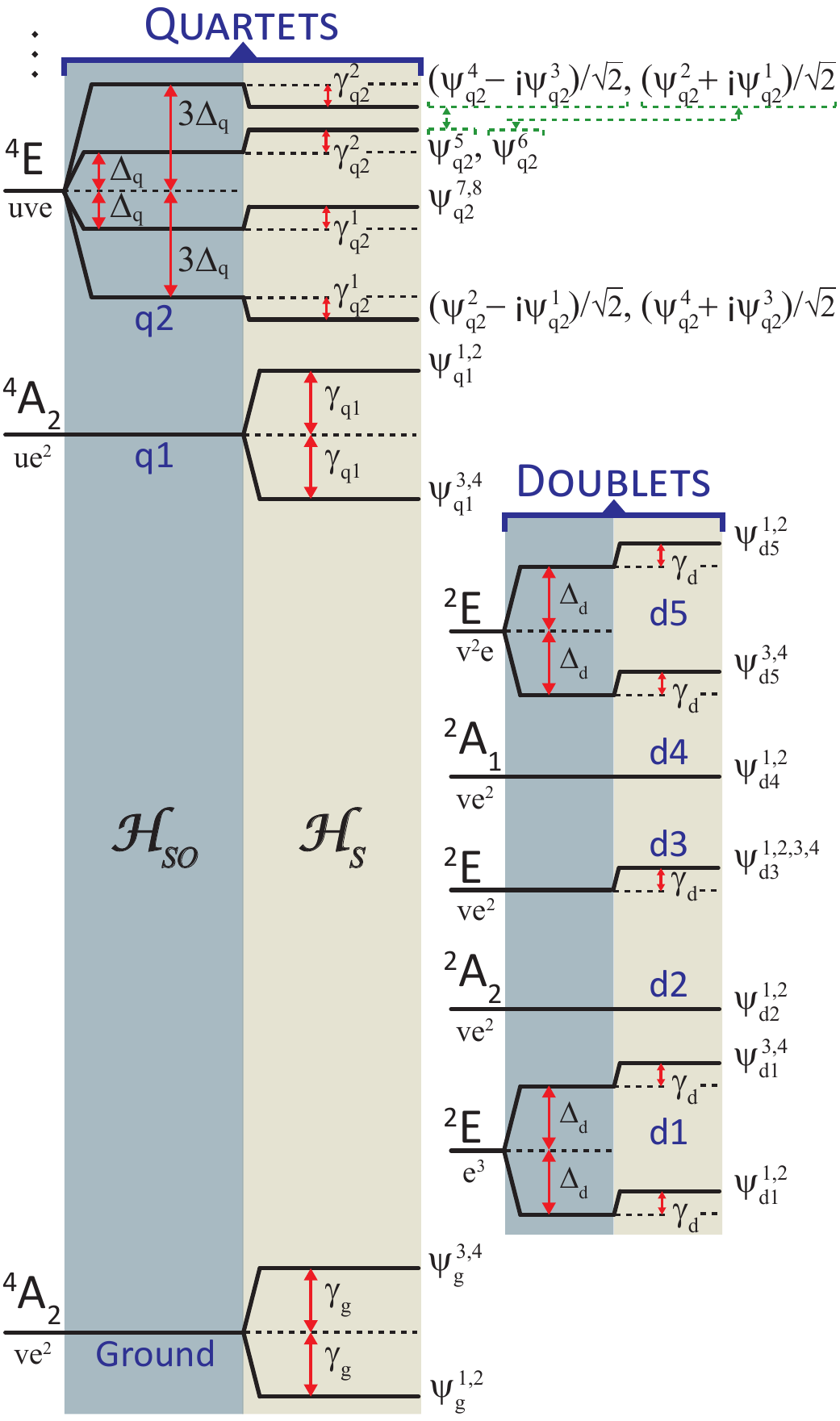}
  \caption{(color online) Electronic configuration of $\textrm{V}_{\textrm{Si}}^-$, shown in terms of the wave functions given in Table\,\ref{Table3}. The splittings are shown explicitly for the SO and spin-spin interactions. The spin quartets are grouped on the left half whereas the metastable doublets are on the right. The states with subscript q and d denote excited quartet and doublet states, respectively. The dashed (green) arrows indicate the mixing due to spin-spin interactions.
 }
  \label{Fig2}
\end{figure}

\begin{figure}[ht]
  \centering
  \includegraphics[width=8.5 cm]{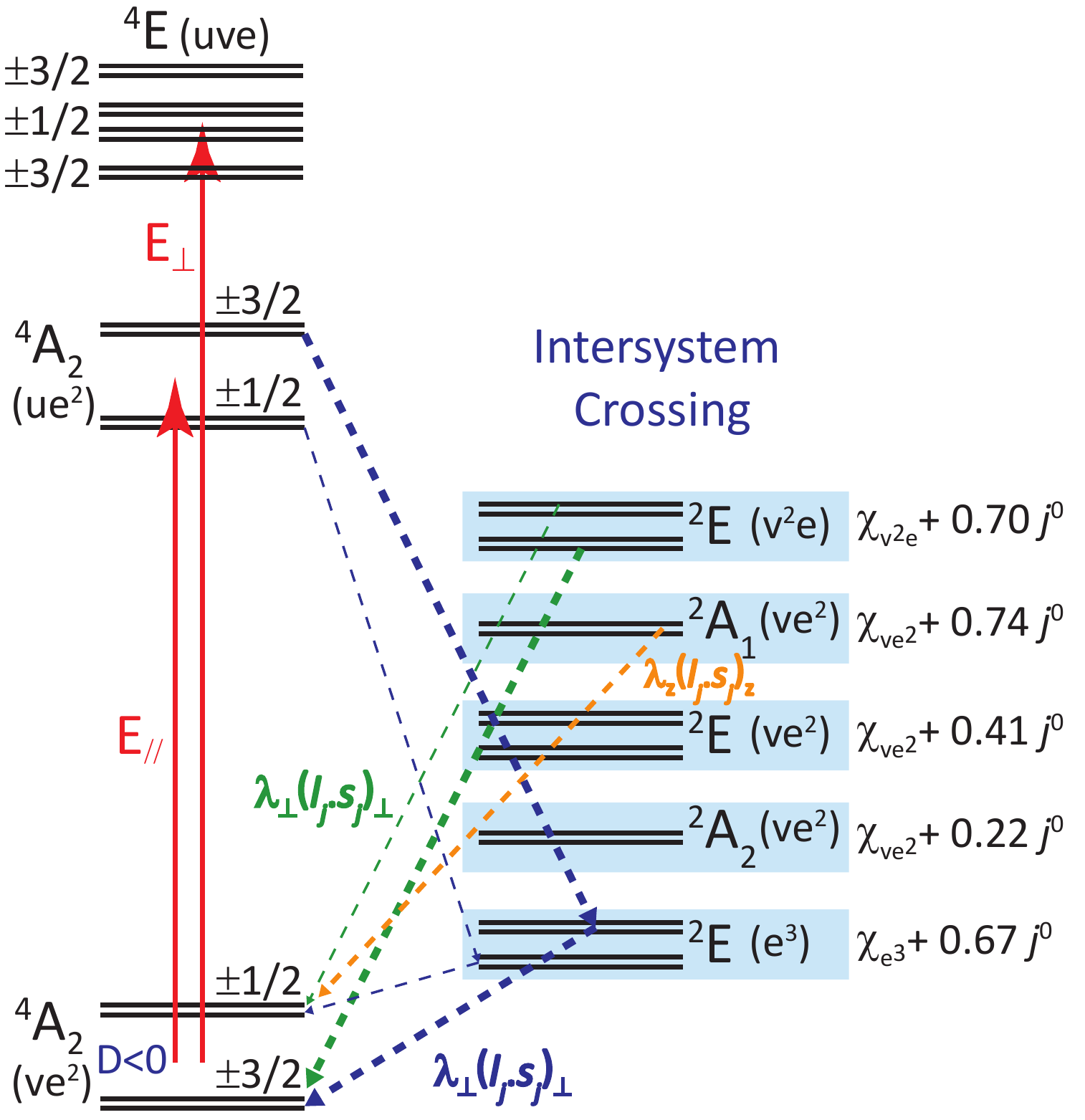}
	\caption{(color online) Spin polarization channel of $\textrm{V}_{\textrm{Si}}^-$ through the spin-orbit assisted dominant intersystem crossing $\prescript{4}{}A_{2}(ue^2){\rightarrow}\prescript{2}{}E(e^3){\rightarrow}\prescript{4}{}A_{2}(ve^2)$ and all other allowed channels are shown in dashed lines. Thicker lines of blue and green indicate 3$\times$ faster transition rate from or to $m_s=\pm 3/2$ states by the transverse component of spin-orbit $\lambda_\bot$ whereas orange represents a channel via the longitudinal $\lambda_z$ component. Energies of the doublets are ordered in terms of the one-particle Coulomb Hamiltonian $\chi=\langle\phi|\sum h_i|\phi\rangle$ and leading many-particle direct integrals, i.e. $j^0=\int\rho_{aa}(r_1) V_{R}(r_1,r_2)\rho_{aa}(r_2)d^3r_1 d^3r_2$, of Coulomb repulsion \cite{supplement}.
 }
  \label{Fig3}
\end{figure}

The states are split and mixed further by spin-orbit (SO) and spin-spin interactions. The SO coupling is
\begin{equation}
\mathcal{H}_{SO}=\sum_j \bm{\ell}_j\cdot \bm{s}_j,\label{3}
\end{equation}
where $\ell_j$ and $s_j$ are orbital and spin angular momentum operators belonging to the $j^{\mathrm{th}}$ hole. The former is defined as $(\ell_j)_i=\epsilon_{ikl}[\nabla V(\bm{r}_j)]_k[\bm{p}_j]_l/2m^2 c^2$ where the $V(\bm{r_j})$ is the local potential, $\bm{p_j}$ is the hole momentum operator with coordinate indices $i,k,l$. The components of both $\mathbf{\ell}$ and $\mathbf{s}$  transform as the $(E_Y,E_X,A_2)$ representation and the $H_{SO}$ Hamiltonian itself transforms as $A_1$. With these symmetry classifications we see that the diagonal part of $H_{SO}$, $\sum_j \ell_{j,z} s_{j,z}$, will only couple states of the same $L$ and $S$ and of orbital symmetry $E$ (since $A_1{\subset}E{\otimes} A_2{\otimes}E$). Thus, the ground states do not split due to this term, while states $\{\Psi^j_\textrm{d}\}$ and $\{\Psi^j_\textrm{q2}\}$ shift and/or mix within their manifolds, as shown in Fig.\,\ref{Fig2} by $\Delta_{\textrm{d}}{=}\langle\phi^E_\xi||L_z^{A_2}||\phi^E_\xi\rangle/(2\sqrt{2})$ and $\Delta_{\textrm{q}}{=}\langle\phi^E_{uve}||L_z^{A_2}||\phi^E_{uve}\rangle/(2\sqrt{2})$  respectively (given in terms of reduced matrix elements and $\xi=\{e^3, v^2e\}$). Note that the total orbital angular momentum operator is used here, which is equivalent to using Eq. \ref{3} for matrix elements between states of the same total $S$ and $L$ \cite{Tinkham2003}.

The transverse parts of the SO interaction, $\sum_j \ell_{j,\bot} s_{j,\bot}$, couple states of different total spin and orbital character $\{u,v\}$ to both $e_x$ and $e_y$ at single particle level. Hence the ground states will couple to $\{\Psi^j_\textrm{d1}\}$ (defined in Table\,\ref{Table3}) via these transverse SO terms. This coupling is crucial both in explaining existing experiments and in designing future applications. The key is to notice that ground states and q1 excited states with $|S_z|{=}3/2$ couple more strongly to excited $\{\Psi^j_\textrm{d1}\}$ ($e^3$) states compared to the states with $|S_z|{=}1/2$. In fact using the states of Table\,\ref{Table3} we can show that the ratio of the matrix elements is $\sqrt{3}$. From this we identify the dominant intersystem crossing channel that constitutes the spin polarization mechanism seen in recent experiments at the single-spin level \cite{Widmann_nmat15} with h-site ($V_2$) defects, where optical driving polarizes the system into the $|S_z|{=}3/2$ states. This mechanism, shown in Fig.\,\ref{Fig3}, also successfully predicts the recently seen increase in the ODMR photo-luminescence intensity with microwave drive \cite{Sorman_PRB2000,Baranov_prb11,Kraus_NP14,Widmann_nmat15,Carter_PRBRC15}.

We can also consider first-order perturbing corrections to the ground state wave functions from the excited dark doublet states through spin-orbit coupling (see Fig.\,\ref{Fig3}). The different strength of the SO matrix elements (e.g., the extra involvement of $l_{j,z}s_{j,z}$ with $m_s=\pm 1/2$ states only) will cause a different degree of admixture of excited states to the $|S_z|{=}3/2$ and $|S_z|{=}1/2$ ground states, which in turn will allow an electric field \cite{Dolde_NP11}, strain and mechanical motion \cite{Soykal_PRL11,Grinolds_NanoLett12,MacQuarrie_PRL13} to couple ground states with different $|S_z|$ projections. This paves the way toward unexplored SiC-based applications in sensing.

Next we consider the spin-spin interaction between the holes. The Hamiltonian is
\begin{equation}
\mathcal{H}_{S}=\frac{\mu_0 g^2 \mu_B^2}{4\pi}\sum_{i>j}\frac{1}{r_{ij}^3}\left\{\bm{s}_i\cdot \bm{s}_j-3\left(\bm{s}_i\cdot \bm{\hat{r}}_{ij}\right)\left(\bm{s}_j\cdot \bm{\hat{r}}_{ij}\right)\right\},\label{4}
\end{equation}
where $g$ is the electron g-factor, $\mu_0$ is the vacuum permeability, and $\mu_B$ is the Bohr magneton. The spin operator of each hole, the distance to each other and its unit vector are $s_{i}$, $r_{ij}$ and $\bm{\hat{r}}_{ij}$, respectively. The spin-spin splittings of the quartets and doublets are shown in Fig.\,\ref{Fig2} in terms of the splitting parameters defined as $\gamma_\textrm{g}{=}\gamma_0 \langle\phi^{A_2}_{ve^2}||I_2||\phi^{A_2}_{ve^2}\rangle/\sqrt{10}$, $\gamma_{\textrm{q1}}{=}\gamma_0 \langle\phi^{A_2}_{ue^2}||I_2||\phi^{A_2}_{ue^2}\rangle/\sqrt{10}$, $\gamma_\textrm{d}{=}\gamma_0\langle\phi^{E}_{\xi}||I_2||\phi^{E}_{\xi}\rangle/(6\sqrt{10})$, $\gamma_{\textrm{q2}}^1{=}\gamma_0\langle\phi^{E}_{uve}||I_2||\phi^{E}_{uve}\rangle/\allowbreak(2\sqrt{10})$ and $\gamma_{\textrm{q2}}^2{=}\gamma_{\textrm{q2}}^1(1-1.028\zeta)$, where $I_2$ is an irregular solid harmonic of second rank, i.e. $I_l^m{=}\sqrt{4\pi/(2l+1)}Y_l^m/r^{l+1}$, $\gamma_0 {=} \mu_0g^2\mu_B^2/4\pi$, and $\zeta{=}\langle\phi^{E}_{uve}||I_2||\phi^{E}_{uve}\rangle/\Delta_{\textrm{q}}{\approx} 0$, see \cite{supplement}. Using in these expressions the calculated bond lengths $d{=}2.058$ \AA, $a{=}2.055$ \AA, and $\theta_0{=}35.26^\circ$ from our DFT results, we estimate the zero field splitting (ZFS) to be $2|\gamma_\textrm{g}| {=} 68$ MHz, in good agreement with experiments \cite{Mizuochi2002,Kraus_NP14,Carter_PRBRC15,Sorman_PRB2000}. However, we found a negative $D$ for the ground state, i.e. $\mathcal{H}_S{\simeq}D[S_z^2-S(S+1)/3]$, causing $m_s=\pm 1/2$ to be energetically higher than the $m_s=\pm 3/2$ states contrary to the some assumptions of $D{>}0$ in literature. In the limit of perfect tetrahedral ($T_d$) symmetry, our calculation also leads to a vanishing ZFS (0 MHz) consistent with the lack of any ZFS with $\textrm{V}_{\textrm{Si}}^-$ centers in 3C-SiC.

\begin{figure}[ht]
  \centering
  \includegraphics[width=7. cm]{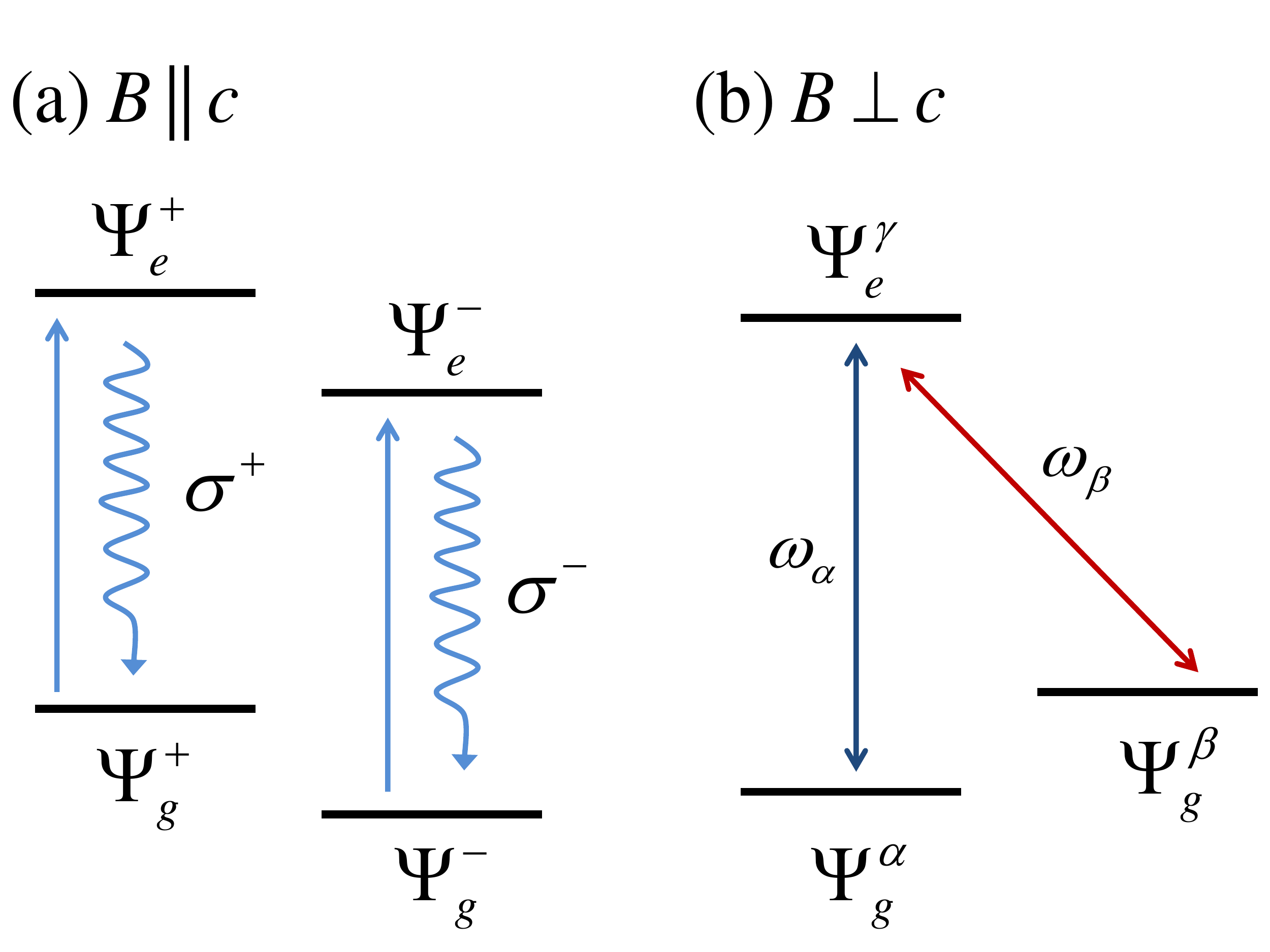}
  \caption{(color online) (a) A B-field parallel to the $C_3$ axis enables the creation of two two-level systems with the same transition frequency but orthogonal polarizations. Periodic coherent pumping followed by spontaneous emission leads to strings of entangled photons. (b) A B-field perpendicular to the $C_3$ axis allows for the creation of a Lambda system.
 }
  \label{Fig4}
\end{figure}

Based on Table\,\ref{Table3}, the rich structure of the various transitions and immunity to all local perturbing electric and strain fields (Kramer's degeneracy) enable the design of a spin-photon interface for applications in quantum computing and quantum communications. Below we propose two such protocols. First consider the ground states with $|S_z|{=}3/2$, split by a B-field along the $C_3$ axis, $\Psi_g^{\pm}{=}\Psi_\textrm{g}^1\pm \Psi_\textrm{g}^2$. The excited states of interest are $\Psi_e^+{=}\Psi^2_\textrm{q2}{-}i\Psi^1_\textrm{q2}$ and $\Psi_e^-{=}\Psi^4_\textrm{q2}{+}i\Psi^3_\textrm{q2}$, which are degenerate energy eigenstates after SO has been included (Fig.\,\ref{Fig2}); these states have $|S_z|{=}3/2$, and since the g-factor is the same in ground and excited states \cite{janzen_physicab09} they split by the same amount as the lower levels. They are also the only states which are not coupled to the states of $\prescript{4}{}A_2$ q1 manifold via $\sum_j \ell_{j,\bot} s_{j,\bot}$ terms. The allowed optical transitions between these sets of states are $\Psi_g^+ {\leftrightarrow} \Psi_e^+$ and $\Psi_g^- {\leftrightarrow} \Psi_e^-$ with right and left circularly polarized light respectively, Fig.\,\ref{Fig4}(a). A coherently excited superposition of  the two excited states decays to an entangled spin-photon state, $|\Psi_g^+\rangle |\sigma +\rangle {+} |\Psi_g^-\rangle |\sigma -\rangle$. Repeating this process produces additional photons, all entangled with the spin and each other, resulting in a multiphoton Greenberger-Horne-Zeilinger state. Augmenting the optical protocol with microwaves can couple the ground states and allow the production of a cluster state \cite{raussendorf_PRL01}, similarly to a proposal for quantum dots \cite{Lindner_PRL09,Economou_PRL10}.

Next we consider a B-field perpendicular to the $C_3$ axis. This mixes all four ground states, and from these we select $\Psi_g^\alpha$ and $\Psi_g^\beta$, along with the excited state $\Psi_e^\gamma$ (all of them given in \cite{supplement} in terms of the states of Table\,\ref{Table3}). Then a $\Lambda$-system  can be formed, Fig.\,\ref{Fig4}(b). This three-level system can be used in numerous quantum applications and demonstrations, including coherent population trapping \cite{santori_optexp06}, optical spin qubit rotations \cite{Economou_PRL07,Yale_PNAS10} and generation of spin-photon entanglement \cite{Togan_science10,Bernien_Nature13} with applications in quantum repeaters \cite{briegel_PRL98}.

In summary, we addressed the crucial need of calculating the multi-particle fine structure of the silicon vacancy defect in SiC. Based on the resulting spectrum we identified the intersystem crossing channel that polarizes the system, found a mechanism to enable quantum sensing applications, and proposed two spin-photon interface protocols. Our work opens further opportunities in understanding these defects and in implementing novel quantum technological applications.

\begin{acknowledgments}
This work was supported in part by ONR. Computer resources were provided by the DoD HPCMP. \"{O}.O.S. and P.D. acknowledge the NRL-NRC Research Associateship Program. We thank S. Carter, Sang-Yun Lee, and Amrit De for comments on the manuscript.
\end{acknowledgments}


\end{document}


\title{Supplementary: Silicon vacancy center in 4H-SiC: Electronic structure and spin-photon interfaces}
\author{\"O. O. Soykal}
\affiliation{NRC post doc residing at Code 6877 Naval Research Laboratory, Washington, D.C. 20375, USA}
\author{Pratibha Dev}
\affiliation{NRC post doc residing at Code 6877 Naval Research Laboratory, Washington, D.C. 20375, USA}
\affiliation{Department of Physics and Astronomy, Howard
University, Washington, DC, USA}
\author{Sophia E. Economou}
\affiliation{Naval Research Laboratory, Washington, DC 20375}
\affiliation{Department of Physics, Virginia Tech, Blacksburg, Virginia 24061, USA}

\maketitle

\section{Single particle molecular orbitals}

\subsection{General remarks}


Unlike the case of NV center in diamond, and other similar defects such as the axial divacancy in SiC, although we expect the single charged Si vacancy to still have a $C_{3\nu}$ symmetry, it can still be interpreted as a very weakly broken $T_d$ symmetry. This is because all four nearest neighbors to the vacancy are carbon atoms with very similar distances (differing by ${\sim}1\%$) from the nearest silicon atom in a perfect crystal, and we do not expect this qualitative feature to change much upon removal of the silicon atom. We thus anticipate that the form of $A_1$ symmetry molecular orbitals will be very close to those of $T_d$:

\begin{align}
\begin{split}
u^{T_d} &= a+b+c+d,\\
v^{T_d} &= a+b+c-3d,
\end{split}\label{1}
\end{align}
where the normalization has been omitted for brevity. Based on the `nearly-$T_d$' symmetry we also anticipate that state $v$ will be near-degenerate with states $e$ (in $T_d$ they are degenerate). As we show below, our single-particle molecular orbitals obtained using DFT indeed confirm these qualitative expectations. Combining results from DFT with analytical calculations we can derive further information, such as the coefficients in the molecular orbitals, overlap integrals and on-site Coulomb energies.

\subsection{First principles calculations}

In order to complement the main group theoretic results, density functional theory (DFT) was used to obtain single particle molecular orbitals (MOs) of the charged Si-vacancy center in 4H-SiC. The ordering of the defect states is obtained from the calculated Kohn-Sham eigenstates around the bandgap of 4H-SiC. The spin-polarized calculations were carried out using the Quantum-ESPRESSO package~\cite{QE-2009}, within the generalized gradient approximation (GGA)~\cite{GGA} of Perdew-Burke-Ernzerhof (PBE)~\cite{PBE}. In this work, we report the results for the V$_{\mathrm{Si}}^{-1}$ at the h-site in a $6\times6\times2$ (576-atoms) supercell with $\Gamma$-centered $2\times2\times2$ k-point sampling according to Monkhort-Pack method.

\begin{figure}[ht]
  \centering
	\subfigure[\,$\bar{u}$ ($A_1$-symmetry)]{\includegraphics[width=3.5 cm]{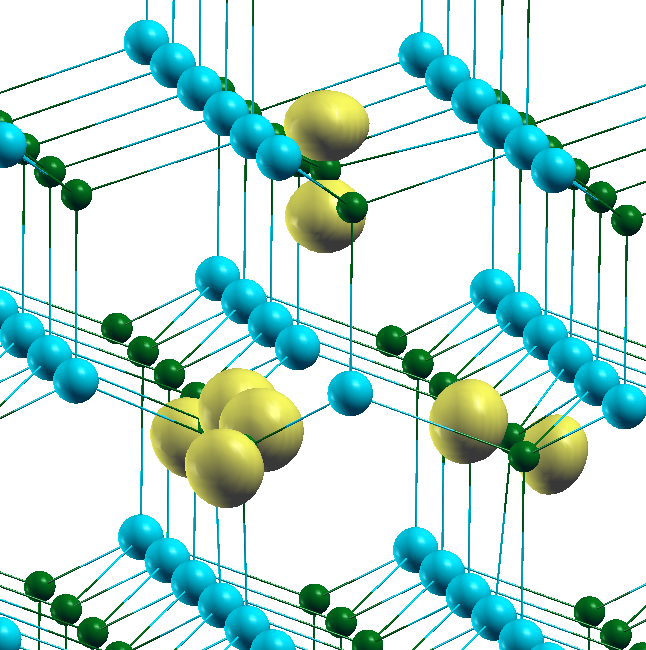}}\hspace{1cm}
	\subfigure[\,$\bar{v}$ ($A_1$-symmetry) ]{\includegraphics[width=3.5 cm]{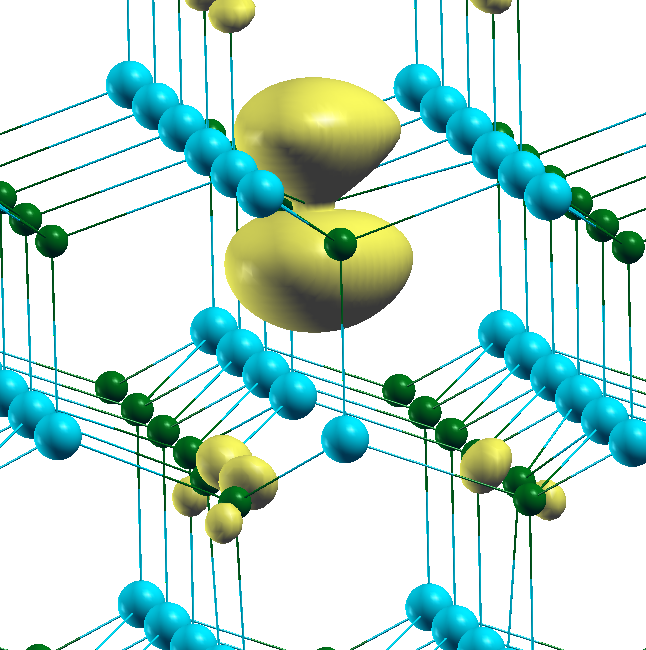}}  \\
	\subfigure[\,$\bar{e}_{x,y}$ ($E$-symmetry) ]{\includegraphics[width=3.5 cm]{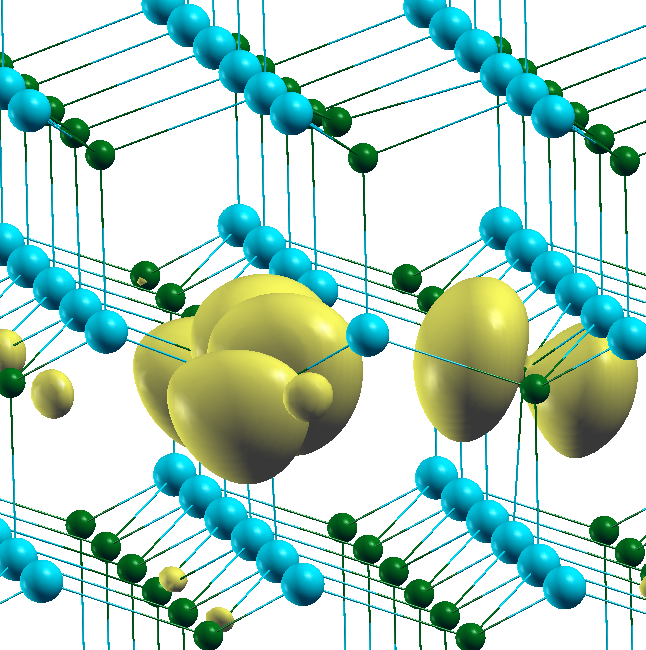}}
	
  \caption{(color online) Isosurface plots ($5\times10^{-3} e/a.u^{-3}$) for the optically-active minority spin MOs of the negatively charged silicon vacancy center $\textrm{V}_{\textrm{Si}}^-$ in 4H-SiC: (a) the highest occupied orbital, $\bar{u}$, (b) the lowest unoccupied orbital, $\bar{v}$, and (c) the next higher unoccupied orbital, $\bar{e}_{x,y}$.}
  \label{Fig1_suppl}
\end{figure}

\begin{figure}[ht]
  \centering
  \includegraphics[width=7.5 cm]{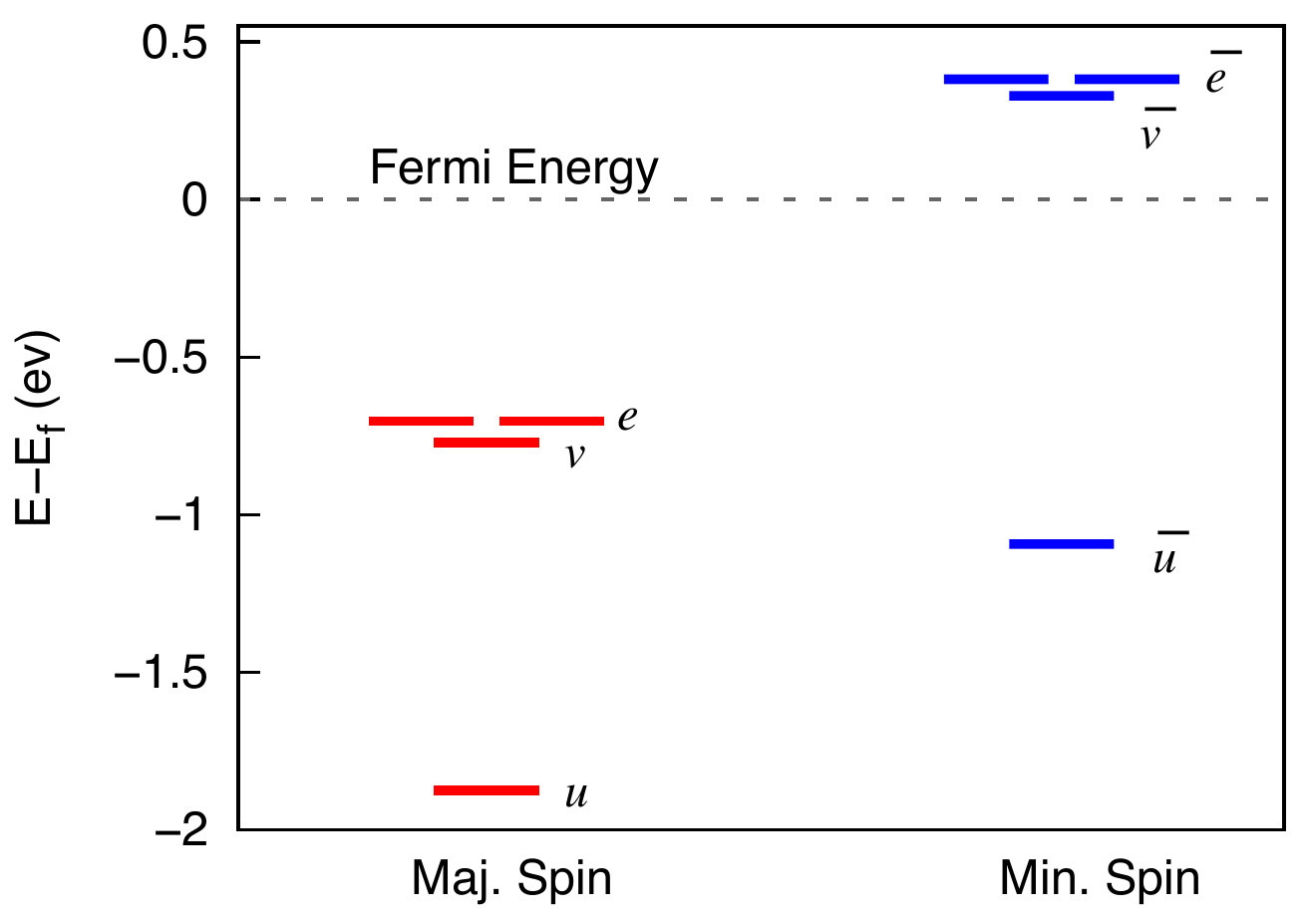}
  \caption{The energy ordering of the defect-induced majority- and minority-spin states.}
    \label{Fig2_suppl}
\end{figure}

The large size of the supercell considered here ensures a reduction in the defect-defect interactions. This produces nearly-flat defect states that are labeled as $u$/$\bar{u}$ ($A_1$-symmetry), $v$/$\bar{v}$ ($A_1$-symmetry)and $e$/$\bar{e}$ ($E$-symmetry). Here, the letters with bar overhead represent the minority spin state, with the excess of three-electrons in the majority spin states. The MOs of the defect plotted in Figure~\ref{Fig1_suppl} differ from those obtained with group theoretic methods (using symmetry-adapted $sp^3$-orbitals) in that they are not restricted to the dangling bonds only. In DFT-calculations no such restriction is made and it includes contributions from other electronic states of the crystal as well. Nonetheless, the defect states can be seen to be highly localized on the carbon atoms surrounding the defect. The majority spin $u$, is found to be resonant with the valence band, while the higher energy defect-states lie in the band gap. This ordering of the defect states can be seen in Fig.\ref{Fig2_suppl}. Thus, the DFT-results reproduce the correct symmetries expected from the group theoretic results and provide the ordering of the defect states relative to each other.

\begin{figure}[ht]
  \centering
	\subfigure[\,$\bar{u}$ ($A_1$-symmetry)]{\includegraphics[width=3.5 cm]{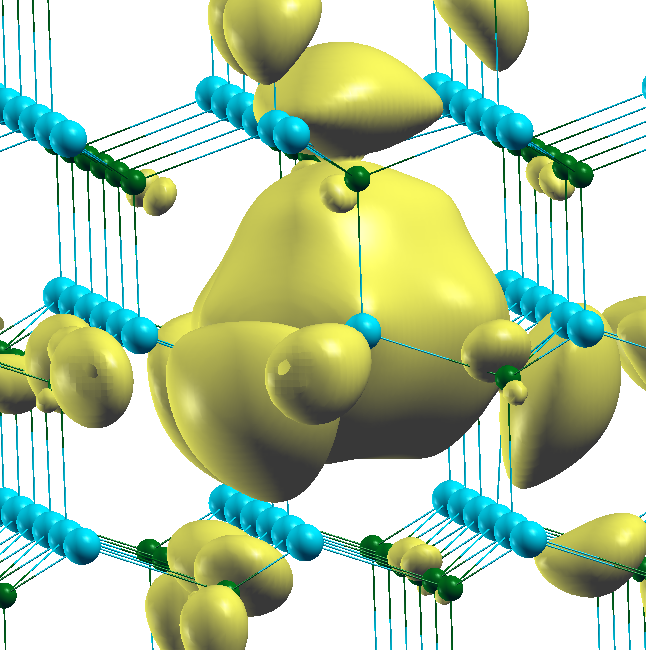}}\hspace{1cm}
	\subfigure[\,$\bar{v}$ ($A_1$-symmetry) ]{\includegraphics[width=3.5 cm]{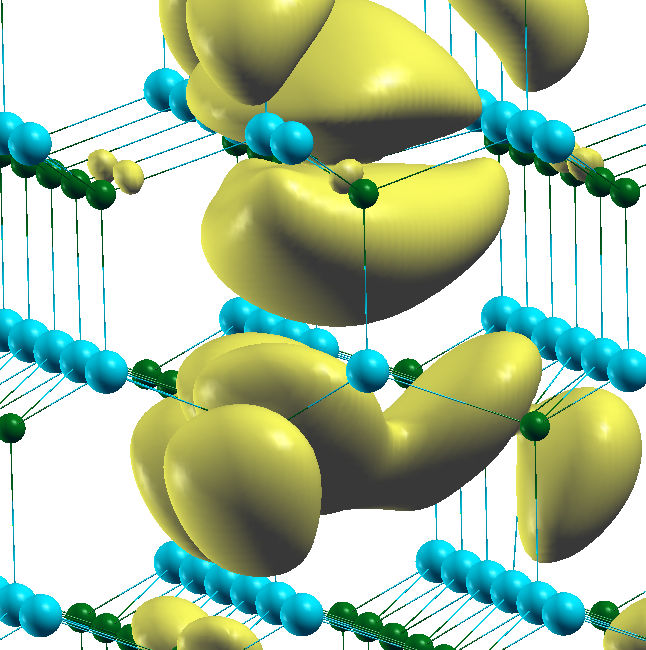}}  \\
  \caption{(color online) Isosurface plots ($5\times10^{-4} e/a.u^{-3}$) for the optically-active minority spin MOs with $A_1$-symmetry, showing: (a) bonding character of $u$, and (b) anti-bonding character of $v$.}
  \label{Fig3_suppl}
\end{figure}

In the main text we used group-theoretic approach to obtain single-particle MOs from the symmetry-adapted linear combinations of the $sp^3$-orbitals belonging to the four carbons surrounding the silicon vacancy. However, group theory does not yield the relative ordering of states with same symmetry, which can be obtained from the DFT calculations. In Fig.\ref{Fig3_suppl}, we choose a different isosurface (compared to the isosurface plots in Fig.\ref{Fig1_suppl}) to showcase the bonding- and the anti-bonding characters of the $A_1$ symmetry states $u$ and $v$, respectively. Thus, DFT results can be used to shed light on the relative ordering of the MOs qualitatively (bonding vs. anti-bonding) and quantitatively (Fig.\ref{Fig2_suppl}).

\subsection{Coulomb interaction and overlap integrals}

The Coulomb interaction Hamiltonian can be grouped as $V_c=\sum_{i\neq j}v_{ij}+\sum_i v_{ii}$ in terms of interactions between different sites (denoted by $ij$) and on-site ($ii$) interactions. Therefore, the Schr\"{o}dinger equation in the basis of $sp^3$ dangling bonds \cite{Huckel31} takes the form of
\begin{equation}
\left(\begin{array}{cccc}
v_{aa} & v_{ab} & v_{ab} & v_{ad}\\
v_{ab} & v_{aa} & v_{ab} & v_{ad}\\
v_{ab} & v_{ab} & v_{aa} & v_{ad}\\
v_{ab} & v_{ab} &v_{ab} & v_{dd}
\end{array}\right)=E_n\left(\begin{array}{cccc}
1 & \lambda_1 & \lambda_1 & \lambda_2\\
\lambda_1 & 1 & \lambda_1 & \lambda_2\\
\lambda_1 & \lambda_1 & 1 & \lambda_2\\
\lambda_2 & \lambda_2 & \lambda_2 & 1
\end{array}\right)\label{2}
\end{equation}
in terms of the overlap integrals $\lambda_1=\int \psi_a\psi_b\,d^3r$ and $\lambda_2=\int \psi_a\psi_d\,d^3r$ between the bonds. For Eq.~(\ref{2}) to have non-trivial solutions for each eigenenergy $E_n$ ($n=u,v,e_x,e_y$), the following determinant has to be zero,
\begin{equation}
\begin{vmatrix}
v_{aa}-E_n & v_{ab}-E_n\lambda_1 & v_{ab}-E_n\lambda_1 & v_{ad}-E_n\lambda_2\\
v_{ab}-E_n\lambda_1 & v_{aa}-E_n & v_{ab}-E_n\lambda_1 & v_{ad}-E_n\lambda_2\\
v_{ab}-E_n\lambda_1 & v_{ab}-E_n\lambda_1 & v_{aa}-E_n & v_{ad}-E_n\lambda_2\\
v_{ab}-E_n\lambda_2 & v_{ab}-E_n\lambda_2 &v_{ab}-E_n\lambda_2 & v_{dd}-E_n
\end{vmatrix}=0\label{3}
\end{equation}
Note that the sites $a,b,c$ are equivalent due to the symmetry of the basal plane. The Coulomb interaction between the sites $a$ and $d$ is roughly equal to that of between $a$ and $b$, i.e. $|v_{ad}|=(1-\delta) |v_{ab}|$ where $\delta\gtrapprox 0$, since the bond length along the c-axis is only slightly distorted from the basal ones as shown by the density functional theory calculations. Since all sites $a$-$d$ have carbon atoms, $|v_{dd}|=|v_{aa}|=v_0$. Moreover, the off-site Coulomb interactions are smaller than the on-site interactions, because of the $1/r$ dependence of the electrostatic potentials, which can be expressed as $|v_{ab}|=\epsilon |v_{aa}|$.

Solutions of Eq.~(\ref{3}), with the realistic assumption $\lim{\delta\to 0}$, leads to the energies of MOs:
\begin{align}
E_u=&-\frac{v_0(\kappa+\Delta\kappa)}{1+2\lambda_1-3\lambda_2^2},\nonumber\\
E_v=&-\frac{v_0(\kappa-\Delta\kappa)}{1+2\lambda_1-3\lambda_2^2},\label{4}\\
E_{e_{x,y}}=&-\frac{v_0(1-\epsilon)}{1-\lambda_1},\nonumber
\end{align}
up to $O(\epsilon^2)$. The coefficients $\kappa$ and $\Delta\kappa$ are given by
\begin{align}\begin{split}
\kappa=&1+\lambda_1+\epsilon(1-3\lambda_2),\\
\Delta\kappa=&\left[\lambda_1^2+3\lambda_2^2+\epsilon^2 (6\lambda_1-6\lambda_2+4)\right.\\
&\left.+\epsilon\left(-6\lambda_1\lambda_2-2\lambda_1+6\lambda_2^2-6\lambda_2\right)\right]^{1/2},
\end{split}\label{5}\end{align}
in terms of the overlap integrals $\lambda_1$, $\lambda_2$, and the off-site to on-site Coulomb ratio $\epsilon$. In the case of zero overlap between the bonds ($\lambda_1=\lambda_2=0$), according to Eq.~(\ref{4}), the energies $E_v$ and $E_{e_{x,y}}$ become equal, i.e. $E_v=E_{e_{x,y}}=-v_0(1-\epsilon)$ and $E_u=-v_0(1+3\epsilon)$, also indicating that $E_u<E_v$. This can be understood as the defect's asymptotic limit to tetrahedral symmetry.

The true benefit of the above treatment is realized once it is used in conjunction with the energies calculated by DFT. By using the MO energies obtained by DFT (Fig.\ref{Fig2_suppl}) in Eq.~(\ref{4}), we find the previously unknown overlap integrals, the on-site potential energy and the Coulomb ratio of the defect to be $\lambda_1=0.0034$, $\lambda_2=0.054$, $v_0=1.177$eV, and $\epsilon=0.285$, respectively. Furthermore, we calculate the eigenfunctions satisfying Eq.~(\ref{3}) as,
\begin{align}\begin{split}
u &= \alpha_u (a+b+c)+\beta_u d,\\
v &= \alpha_v (a+b+c)+\beta_v d,\\
e_x &= \alpha_x (2c-a-b),\\
e_y &= \alpha_y (a-b),
\end{split}\label{6}\end{align}
with the coefficients obtained as $\alpha_u=0.523$, $\beta_u=0.423$, $\alpha_v=-0.272$, $\beta_v=0.882$, $\alpha_x=0.408$, and $\alpha_y=0.707$. The coefficients of $u$ and $v$ only slightly differ from the readily known coefficients of $T_d$ symmetry \cite{Tinkham2003}, i.e. $\alpha_u=\beta_u=0.5$, $\alpha_v=-0.289$, and $\beta_v=0.866$.  Later on, we use these coefficients to estimate the zero-field splitting of the ground state leading to a remarkable agreement with the experimentally measured values.

\subsection{Energy order of the doublets}

Due to the many-particle nature of the doublets, we cannot obtain the ordering of the states using DFT, which is an effective single-particle description of the system. Therefore, we analyzed the ordering via Coulomb Hamiltonian $H_c{=}\sum h_i+\sum_{i,j}V_{ee}(r_i,r_j)$ using the wave functions of the states $\Psi^i_{\textrm{d1-d5}}$ given in Table. I of the main text. One electron (hole) Coulomb terms are included in $h_i$, whereas $V_{ee}(r_i,r_j)=e^2/(4\pi\epsilon_0 |r_i-r_j|)$ is the two-particle Coulomb repulsion potential. Eigen values of $h_i$ in MOs basis are represented by $\chi$ and can be estimated from DFT. Many-particle Coulomb integrals are given as $j^0_{ll}{=}{\int}\rho_{ll}(1)V_{ee}\rho_{ll}(2)d^3r_1 d^3r_2$, $j_{lm}{=}{\int}\rho_{ll}(1)V_{ee}\rho_{mm}(2)d^3r_1 d^3r_2$, and $k_{lm}{=}{\int}\rho_{lm}(1)V_{ee}\rho_{lm}(2)d^3r_1 d^3r_2$. The integrals $j^0_{ll}$, $j_{lm}$, and $k_{lm}$ are the one-center Coulomb integral, two-particle Coulomb repulsion direct and exchange integrals, respectively. Charge density is defined as $\rho_{lm}(i)=\psi_l(i)^*\psi_m(i)$ belonging to the $i^\textrm{th}$ particle in the basis of sp$^3$ hybridized dangling bond wave functions with $l,m{=}\{a,b,c,d\}$ and $l{\neq}m$. We obtain the Coulomb energies of doublets as,
\begin{align}
\textrm{E}^{E}_{e^3}=&\chi_{e^3}+0.67 j^0_{aa}+2.33 j_{ab}-0.33 k_{ab}\label{7}\\
\textrm{E}^{A_2}_{ve^2}=&\chi_{ve^2}+0.22 j^0_{aa}+1.22 j_{ab}+1.56 j_{ad}\nonumber\\
&-1.22 k_{ab}+0.78 k_{ad}\label{8}\\
\textrm{E}^{E}_{ve^2}=&\chi_{ve^2}+0.41 j^0_{aa}+1.04 j_{ab}+1.56 j_{ad}\nonumber\\
&-0.04 k_{ab}-0.78 k_{ad}\label{9}\\
\textrm{E}^{A_1}_{ve^2}=&\chi_{ve^2}+0.74 j^0_{aa}+0.70 j_{ab}+1.56 j_{ad}\nonumber\\
&+1.30 k_{ab}-0.78 k_{ad}\label{10}\\
\textrm{E}^{E}_{v^2e}=&\chi_{v^2e}+0.09 j^0_{aa}+0.61 j^0_{dd}+0.40 j_{ab}+1.90 j_{ad}\nonumber\\
&-0.30 k_{ab}-0.89 k_{ad}\label{11}
\end{align}
where the relationship $\chi\gg j^0\gg j \gg k$ holds and due to the nearly $T_d$ symmetry of the center charge localization on the basal and z-axis carbon atoms are assumed to be similar, i.e., $j^0_{aa}\simeq j^0_{dd}$. Furthermore, we obtain the ground state energy in a similar way:
\begin{equation}
\textrm{E}_g=1.44 (j_{ab}-k_{ab})+1.56(j_{ad}-k_{ad}).\label{12}
\end{equation}
Assuming $\chi_{v^2e}>\chi_{ve^2}>\chi_{e^3}$, the ordering of doublets becomes $\textrm{E}^{E}_{e^3}$, $\textrm{E}^{A_2}_{ve^2}$, $\textrm{E}^{E}_{ve^2}$, $\textrm{E}^{A_1}_{ve^2}$, and $\textrm{E}^E_{v^2e}$ increasing in energy.

\section{Three-particle states}

Because of the near-degeneracy of state $v$ with states $e_x$ and $e_y$, it is energetically favorable for two electrons to occupy the $e$ states instead of paying the energetic cost of doubly occupying the only slightly lower in energy state $v$. As a result, in the ground state the occupied states are $v$, $e_x$ and $e_y$, in the three-hole picture.

For the $ve^2$ ground state, the three hole configuration space is spanned by $32$ ($2\otimes 4\otimes 4$) basis functions in the form of single particle Kronecker products ${f_\kappa^j}=(\{v\}\otimes\{\alpha,\beta\})\otimes(\{e_x,e_y\}\otimes\{\alpha,\beta\})\otimes(\{e_x,e_y\}\otimes\{\alpha,\beta\})$. However, consideration of the Pauli exclusion principle discards $8$ of them leaving $24$ basis states.

Moreover, the single particle irreducible matrix representations for the cases where the degeneracy lies only in the orbital, only in the spin, or in both spaces are simply $\Gamma_{E}(R)\otimes\mathbb{1}_s$, $\mathbb{1}_o\otimes\Gamma_{E_{1/2}}(R)$, or $\Gamma_{E}(R)\otimes\Gamma_{E_{1/2}}(R)$, respectively. Note that the identity matrices are defined as $\mathbb{1}_o$ for the orbital and $\mathbb{1}_s$ for the spin subspace. The explicit form of the matrices $\Gamma(R)$  are given in Table \ref{St1}.

\setlength{\tabcolsep}{5pt}
\renewcommand{\arraystretch}{1.2}
\begin{table}[ht]
\centering
\begin{tabular}{c|c|c}
$R$ & $\Gamma_E (R)$ & $\Gamma_{E_{1/2}}(R)$\\
\hline\hline
$\left\{\begin{array}{c} E\\ \bar{E}\end{array}\right\}$ & $\left[\begin{array}{cc} 1 & 0 \\ 0 & 1 \end{array}\right]$ & $\pm\left[\begin{array}{cc} 1 & 0 \\ 0 & 1 \end{array}\right]$\Tstrut\\[12pt]
$\left\{\begin{array}{c} C_3^+ \\ \bar{C}_3^+\end{array}\right\}$ & $\left[\begin{array}{cc} -\frac{1}{2} & -\frac{\sqrt{3}}{2} \\ \frac{\sqrt{3}}{2} & -\frac{1}{2} \end{array}\right]$ & $\pm\left[\begin{array}{cc} \bar{\epsilon} & 0 \\ 0 & \bar{\epsilon}^* \end{array}\right]$\\[12pt]
$\left\{\begin{array}{c} C_3^- \\ \bar{C}_3^-\end{array}\right\}$ & $\left[\begin{array}{cc} -\frac{1}{2} & \frac{\sqrt{3}}{2} \\ -\frac{\sqrt{3}}{2} & -\frac{1}{2} \end{array}\right]$ & $\pm\left[\begin{array}{cc} \bar{\epsilon}^* & 0 \\ 0 & \bar{\epsilon} \end{array}\right]$\\[12pt]
$\left\{\begin{array}{c} \sigma_{\nu 1} \\ \bar{\sigma}_{\nu 1}\end{array}\right\}$ & $\left[\begin{array}{cc} 1 & 0 \\ 0 & -1 \end{array}\right]$ & $\pm\left[\begin{array}{cc} 0 & \bar{1} \\ 1 & 0 \end{array}\right]$\\[12pt]
$\left\{\begin{array}{c} \sigma_{\nu 2} \\ \bar{\sigma}_{\nu 2}\end{array}\right\}$ & $\left[\begin{array}{cc} -\frac{1}{2} & -\frac{\sqrt{3}}{2} \\ -\frac{\sqrt{3}}{2} & \frac{1}{2} \end{array}\right]$ & $\pm\left[\begin{array}{cc} 0 & \bar{\epsilon}^* \\ \epsilon & 0 \end{array}\right]$\\[12pt]
$\left\{\begin{array}{c} \sigma_{\nu 3} \\ \bar{\sigma}_{\nu 3}\end{array}\right\}$ & $\left[\begin{array}{cc} -\frac{1}{2} & \frac{\sqrt{3}}{2} \\ \frac{\sqrt{3}}{2} & \frac{1}{2} \end{array}\right]$ & $\pm\left[\begin{array}{cc} 0 & \bar{\epsilon} \\ \epsilon^* & 0 \end{array}\right]$
\end{tabular}
\caption{Irreducible matrix representations of $E$ and $E_{1/2}$ for orbital and spin degrees of freedom, respectively. $\Gamma_{E_{1/2}}$ is given in helicity basis with $\epsilon=\exp{i2\pi/3}$.}
\label{St1}
\end{table}

For the multi-particle $ve_xe_y$ ground state configuration, the irreducible matrix representation $\Gamma_{\lambda\kappa}^{(j)}(R)$ can be decomposed into its orbital and spin components for each particle, i.e. $\Gamma_{\lambda\kappa}^{(j)}(R)=\left[(\mathbb{1}\otimes\Gamma_{E_{1/2}})\otimes(\Gamma_E\otimes\Gamma_{E_{1/2}})\otimes(\Gamma_E\otimes\Gamma_{E_{1/2}})\right]_{\lambda\kappa}^{(j)}(R)$. In this form, application of the projection operator \cite{Tinkham2003} on each basis function,
\begin{equation}
\mathcal{P}^{(j)}f_\kappa^j=(I_j/h)\sum_R\sum_\lambda^{I_j}\chi^{(j)}(R)^*\Gamma_{\lambda\kappa}^{(j)}(R)f_\lambda^j,\label{13}
\end{equation}
yields the symmetry adapted basis functions belonging to the $j^{\textrm{th}}$ representation of the ground state. Character table of $C_{3\nu}$ is given in Table \ref{Table1}. This gives us a prescription for generating all the partners of any basis function belonging to a given representation. Further combinations of these symmetry adapted basis functions are then formed according to the spin configurations listed in Table \ref{St2} to finally obtain all the quartet and doublet wave functions of $ve^2$ configuration listed in Table II of the main text. The wave functions for the $uve$ excited state quartet (q2) are also produced in the same way.

\setlength{\tabcolsep}{0.5pt}
\renewcommand{\arraystretch}{1.2}
\begin{table}[!hb]
\centering
\begin{tabular}{|c|C{0.5cm}|C{0.5cm}|C{0.5cm}|C{0.5cm}|C{0.5cm}|C{0.5cm}|C{0.5cm}|C{0.5cm}|C{0.5cm}|C{0.5cm}|C{0.5cm}|C{0.5cm}|}
\hline
 & $E$ & $C_3^+$ & $C_3^-$ & $\sigma_{\nu 1}$ & $\sigma_{\nu 2}$ & $\sigma_{\nu 3}$  & $\bar{E}$ & ${\bar{C}_3^+}$ & $\bar{C}_3^-$ & $\bar{\sigma}_{\nu 1}$ & $\bar{\sigma}_{\nu 2}$ & $\bar{\sigma}_{\nu 3}$\\
\hline
$A_1$ & 1 & 1 & 1 & 1 & 1 & 1 & 1 & 1 & 1 & 1 & 1 & 1 \\ \rowcolor{gray!10}
$A_2$ & 1 & 1 & 1 & -1 & -1 & -1 & 1 & 1 & 1 & -1 & -1 & -1\\
$E$ & 2 & 1 & 1 & 0 & 0 & 0 & 2 & 1 & 1 & 0 & 0 & 0 \\ \rowcolor{gray!10}
$E_{1/2}$ & 2 & 1 & 1 & 0 & 0 & 0 & -2 & -1 & -1 & 0 & 0 & 0 \\
$\prescript{1}{}E_{3/2}$ & 1 & -1 & -1 & i & i & i & -1 & 1 & 1 & -i & -i & -i \\ \rowcolor{gray!10}
$\prescript{2}{}E_{3/2}$ & 1 & -1 & -1 & -i & -i & -i & -1 & 1 & 1 & i & i & i \\
\hline
\end{tabular}
\caption{Character table of $C_{3\nu}$ double group. 3-particle coordinate (spin) space belongs to the first (last) three rows.}
\label{Table1}
\end{table}

\setlength{\tabcolsep}{5pt}
\renewcommand{\arraystretch}{1.1}
\begin{table}[ht]
\centering
\begin{tabular}{|c|c|c|l|}
\hline
\multicolumn{4}{|c|}{$D^{1/2}\otimes D^{1/2}\otimes D^{1/2}$} \\
\hline
$\Gamma_s$ & $S$ & $m_s$ & \multicolumn{1}{c|}{$\psi_{S}^{m_s}$} \\
\hline
\multirow{4}{*}{$D^{3/2}$} & \multirow{4}{*}{$3/2$}     & $+3/2$ & $|\alpha\alpha\alpha\rangle$ \\
											     &														& $+1/2$ & $|\alpha\alpha\beta\rangle+|\alpha\beta\alpha\rangle+|\beta\alpha\alpha\rangle$ \\
													 &														& $-1/2$ & $|\beta\beta\alpha\rangle+|\beta\alpha\beta\rangle+|\alpha\beta\beta\rangle$ \\
											     &														& $-3/2$ & $|\beta\beta\beta\rangle$ \\
\hline
\multirow{2}{*}{$D^{1/2}$} & \multirow{2}{*}{$1/2$}			& $+1/2$ & $|\alpha\beta\alpha\rangle-|\beta\alpha\alpha\rangle$ \\
											     &														& $-1/2$ & $|\beta\alpha\beta\rangle-|\alpha\beta\beta\rangle$ \\											
\hline
\multirow{2}{*}{$D^{1/2}$} & \multirow{2}{*}{$1/2$}	 		& $+1/2$ & $|\alpha\beta\alpha\rangle+|\beta\alpha\alpha\rangle-2|\alpha\alpha\beta\rangle$ \\
											     &														& $-1/2$ & $|\beta\alpha\beta\rangle+|\alpha\beta\beta\rangle-2|\beta\beta\alpha\rangle$ \\
\hline
\end{tabular}
\caption{Free space spin configuration of three holes in terms of spin up $\alpha$ and down $\beta$ states reduced into irrep. of a quartet $D^{3/2}$ and two doublets $D^{1/2}$.}
\label{St2}
\end{table}

\section{Spin-orbit assisted transitions amongst dark doublet states}

We show the spin-orbit coupling matrix elements between all doublet manifolds $\textrm{d1}{-}\textrm{d}5$ in Table \ref{St3} using Eq.1 of our manuscript with the symmetry-adapted basis functions given in Table I. The spin-orbit coupling parameters that are perpendicular and parallel to the $C_3$ axis of the defect are represented by $\lambda_\bot$ and $\lambda_{z}$, respectively. Each element of the matrix is evaluated by $\langle \Psi_i || \mathscr{H}_{SO} || \Psi_j \rangle$ where $i$ and $j$ are the wave functions given as the row and column headings. We also omit the dark doublets, much higher in energy, lying in between the excited quartet states q1 and q2. These will either transition to the lowest excited quartet state (q1) or along the doublet ladder to the five lower doublet states. The key thing to notice here is, as shown in Table \ref{St3}, all doublet states except d5 have spin-orbit assisted allowed transitions to the lowest d1 doublet. However, d5 doublet can transition to d1 through the other doublets in between and also has strong transition rate -just like d1- into the ground spin $m_s=\pm 3/2$ states by itself which will assist the optical spin polarization process. Therefore, any other high lying doublet states we omitted in this fine structure will follow the general paths shown in our manuscript and will not affect the dominant spin-polarization channel identified as to be through the d1 doublet in our manuscript. Note that d1 is energetically the closest doublet to the ground state (as shown above) and it is also the only one connected to the q1 quartet with a directly allowed spin-orbit assisted transition.

\setlength{\tabcolsep}{0.5pt}
\renewcommand{\arraystretch}{1.2}
\begin{table*}[!ht]
\centering
\begin{tabular}{|c|C{0.8cm}|C{0.8cm}|C{0.8cm}|C{0.8cm}|C{0.8cm}|C{0.8cm}|C{0.8cm}|C{0.8cm}|C{0.8cm}|C{0.8cm}|C{0.8cm}|C{0.8cm}|C{0.8cm}|C{0.8cm}|C{0.8cm}|C{0.8cm}|}
\hline
 & $\Psi_{\textrm{d}1}^1$ & $\Psi_{\textrm{d}1}^2$ & $\Psi_{\textrm{d}1}^3$ & $\Psi_{\textrm{d}1}^4$ & $\Psi_{\textrm{d}2}^1$ & $\Psi_{\textrm{d}2}^2$ & $\Psi_{\textrm{d}3}^1$ & $\Psi_{\textrm{d}3}^2$ & $\Psi_{\textrm{d}3}^3$ & $\Psi_{\textrm{d}3}^4$ & $\Psi_{\textrm{d}4}^1$ & $\Psi_{\textrm{d}4}^2$ & $\Psi_{\textrm{d}5}^1$ & $\Psi_{\textrm{d}5}^2$ & $\Psi_{\textrm{d}5}^3$ & $\Psi_{\textrm{d}5}^4$ \\
\hline
$\Psi_{\textrm{d}1}^1$ & $-\frac{\lambda_z}{2}$ & 0 & 0 & 0 & 0 & $\frac{-\lambda_\bot}{2\sqrt{3}}$ & 0 & 0 & 0 & 0 & 0 & $\frac{i\lambda_\bot}{2}$ & 0 & 0 & 0 & 0 \\ \rowcolor{gray!10}
$\Psi_{\textrm{d}1}^2$ & 0 & $-\frac{\lambda_z}{2}$ & 0 & 0 & $\frac{-\lambda_\bot}{2\sqrt{3}}$ & 0 & 0 & 0 & 0 & 0 & $\frac{i\lambda_\bot}{2}$ & 0 & 0 & 0 & 0 & 0 \\ 
$\Psi_{\textrm{d}1}^3$ & 0 & 0 & $\frac{\lambda_z}{2}$ & 0 & 0 & 0 & $\frac{-i\lambda_\bot}{2\sqrt{2}}$ & $\frac{-i\lambda_\bot}{2\sqrt{2}}$ & 0 & $\frac{-\lambda_\bot}{2}$ & 0 & 0 & 0 & 0 & 0 & 0 \\ \rowcolor{gray!10}
$\Psi_{\textrm{d}1}^4$ & 0 & 0 & 0 & $\frac{\lambda_z}{2}$ & 0 & 0 & $\frac{i\lambda_\bot}{2\sqrt{2}}$ & $\frac{i\lambda_\bot}{2\sqrt{2}}$ & 0 & $\frac{-\lambda_\bot}{2}$ & 0 & 0 & 0 & 0 & 0 & 0 \\
$\Psi_{\textrm{d}2}^1$ & 0 & $\frac{-\lambda_\bot}{2\sqrt{3}}$ & 0 & 0 & 0 & 0 & 0 & 0 & 0 & 0 & $\frac{i\lambda_z}{\sqrt{3}}$ & 0 & 0 & $\frac{-i\lambda_\bot}{2\sqrt{3}}$ & 0 & 0 \\ \rowcolor{gray!10}
$\Psi_{\textrm{d}2}^2$ & $\frac{-\lambda_\bot}{2\sqrt{3}}$ & 0 & 0 & 0 & 0 & 0 & 0 & 0 & 0 & 0 & 0 & $\frac{i\lambda_z}{\sqrt{3}}$ & $\frac{i\lambda_\bot}{2\sqrt{3}}$ & 0 & 0 & 0 \\
$\Psi_{\textrm{d}3}^1$ & 0 & 0 & $\frac{i\lambda_\bot}{2\sqrt{2}}$ & $\frac{-i\lambda_\bot}{2\sqrt{2}}$ & 0 & 0 & 0 & 0 & 0 & 0 & 0 & 0 & 0 & 0 & $\frac{-\lambda_\bot}{2\sqrt{2}}$ & $\frac{\lambda_\bot}{2\sqrt{2}}$ \\ \rowcolor{gray!10}
$\Psi_{\textrm{d}3}^2$ & 0 & 0 & $\frac{i\lambda_\bot}{2\sqrt{2}}$ & $\frac{-i\lambda_\bot}{2\sqrt{2}}$ & 0 & 0 & 0 & 0 & 0 & 0 & 0 & 0 & 0 & 0 & $\frac{-\lambda_\bot}{2\sqrt{2}}$ & $\frac{\lambda_\bot}{2\sqrt{2}}$ \\
$\Psi_{\textrm{d}3}^3$ & 0 & 0 & 0 & 0 & 0 & 0 & 0 & 0 & 0 & 0 & 0 & 0 & 0 & 0 & 0 & 0 \\ \rowcolor{gray!10}
$\Psi_{\textrm{d}3}^4$ & 0 & 0 & $\frac{-\lambda_\bot}{2}$ & $\frac{-\lambda_\bot}{2}$ & 0 & 0 & 0 & 0 & 0 & 0 & 0 & 0 & 0 & 0 & $\frac{-i\lambda_\bot}{2}$ & $\frac{-i\lambda_\bot}{2}$ \\
$\Psi_{\textrm{d}4}^1$ & 0 & $\frac{-i\lambda_\bot}{2}$ & 0 & 0 & $\frac{-i\lambda_z}{\sqrt{3}}$ & 0 & 0 & 0 & 0 & 0 & 0 & 0 & 0 & $\frac{-\lambda_\bot}{2}$ & 0 & 0 \\ \rowcolor{gray!10}
$\Psi_{\textrm{d}4}^2$ & $\frac{-i\lambda_\bot}{2}$ & 0 & 0 & 0 & 0 & $\frac{-i\lambda_z}{\sqrt{3}}$ & 0 & 0 & 0 & 0 & 0 & 0 & $\frac{\lambda_\bot}{2}$ & 0 & 0 & 0 \\ 
$\Psi_{\textrm{d}5}^1$ & 0 & 0 & 0 & 0 & 0 & $\frac{-i\lambda_\bot}{2\sqrt{3}}$ & 0 & 0 & 0 & 0 & 0 & $\frac{\lambda_\bot}{2}$ & $\frac{\lambda_z}{2}$ & 0 & 0 & 0 \\ \rowcolor{gray!10}
$\Psi_{\textrm{d}5}^2$ & 0 & 0 & 0 & 0 & $\frac{i\lambda_\bot}{2\sqrt{3}}$ & 0 & 0 & 0 & 0 & 0 & $\frac{-\lambda_\bot}{2}$ & 0 & 0 & $\frac{\lambda_z}{2}$ & 0 & 0 \\
$\Psi_{\textrm{d}5}^3$ & 0 & 0 & 0 & 0 & 0 & 0 & $\frac{-\lambda_\bot}{2\sqrt{2}}$ & $\frac{-\lambda_\bot}{2\sqrt{2}}$ & 0 & $\frac{i\lambda_\bot}{2}$ & 0 & 0 & 0 & 0 & $\frac{-\lambda_z}{2}$ & 0 \\ \rowcolor{gray!10}
$\Psi_{\textrm{d}5}^4$ & 0 & 0 & 0 & 0 & 0 & 0 & $\frac{\lambda_\bot}{2\sqrt{2}}$ & $\frac{\lambda_\bot}{2\sqrt{2}}$ & 0 & $\frac{i\lambda_\bot}{2}$ & 0 & 0 & 0 & 0 & 0 & $\frac{-\lambda_z}{2}$\\
\hline
\end{tabular}
\caption{Spin-orbit matrix elements amongst the dark doublet states. Spin-orbit parameters of the defect are given by $\lambda_z$ and $\lambda_\bot$ along the $C_3$ axis and the basal plane of the defect, respectively.}
\label{St3}
\end{table*}

\section{Spin-spin interaction}

\subsection{Spherical tensor components}
The spin dipole-dipole operator given in terms of the single particle operators $S_d=\bm{s}^i\cdot \bm{s}^j-3\left(\bm{s}^i\cdot \bm{\hat{r}}^{ij}\right)\left(\bm{s}^j\cdot \bm{\hat{r}}^{ij}\right)$ can be expressed as $\left\{ A+B+C+D+E+F\right\}$, using the following spherical tensor components,
\begin{align}
&A=-4\sqrt{\pi/5}\; Y_2^0\ s_z^i s_z^j, \quad B=\sqrt{\pi/5}\; Y_2^0\left(s_-^is_+^j+s_+^is_-^j\right),\nonumber\\
&C,D=\mp\sqrt{6\pi/5}\; Y_2^{\mp 1}\left(s_\pm^is_z^j+s_z^is_{\pm}^j\right),\nonumber\\
&E,F=-\sqrt{6\pi/5}\; Y_2^{\mp 2}s_{\pm}^is_{\pm}^j.\label{14}
\end{align}
Orbital parts of $A$ and $B$ terms involving the spherical harmonic $Y_2^0$ belong to the $A_1$ symmetry, whereas all other terms belong to the $E$ symmetry. Since the ground state wave functions (Table II of the main text) possess an $A_2$ orbital symmetry, only $A$ and $B$ terms of Eq.~(\ref{14}) will cause the zero field spin-splitting of the ground state; however, for a q2 excited state with $E$ orbital symmetry and corresponding spin symmetries listed in Table II of the main text, all terms can contribute to the splitting.

We first calculate the matrix elements of $S_d$ for each wave function listed in Table II of the main text by direct evaluation of its spin components. The remaining spatial dependence of the matrix elements can then be analyzed through the Wigner-Eckart theorem using the spatial components of the spherical tensors listed above.

\subsection{Ground state zero-field spin splitting}

In the main text, we report the zero-field spin (ZFS) splitting of the ground state in a compact form as
\begin{equation}
\gamma_g=\gamma_0 \langle\phi^{A_2}_{ve^2}||I_2||\phi^{A_2}_{ve^2}\rangle/\sqrt{10},\label{15}
\end{equation}
where $I_2$ is an irregular solid harmonic of second rank, i.e. $I_l^m=\sqrt{4\pi/(2l+1)}Y_l^m/r^{l+1}$ and $\gamma_0=\mu_0g^2\mu_B^2/(4\pi)$. In its open form, it can be written as
\begin{equation}
\gamma_g=\gamma_0 \sqrt{\frac{\pi}{5}}\sum_{i>j} \langle ve_xe_y||\frac{Y_{2,ij}^0}{r_{ij}^3}||ve_xe_y\rangle,\label{16}
\end{equation}
where $Y_{2,ij}^0/r_{ij}^3$ can be treated as a pair operator. In terms of the direct and exchange integrals \cite{Tinkham2003,Mahan2000}, the expectation value of any pair operator $F$ is given by,
\begin{align}\begin{split}
\langle X||F||X\rangle=\sum_{i>j}&\left\{\langle a_i a_j |f(i,j)|a_i a_j\rangle\right.\\
&\left.-\langle a_i a_j |f(i,j)|a_j a_i\rangle\right\},
\end{split}\label{17}
\end{align}
where $F=\sum_{i>j}f(i,j)$ is the total pair operator and $X$ is the multi-particle antisymmetrized product (Slater determinant) defined as $AP\left[a_1(1)a_2(2)\cdots a_n(N)\right]$. In the case of ground state ZFS, these are $f(i,j)=Y_{2,ij}^0/r_{ij}^3$ and $X=AP\left[v(1)e_x(2)e_y(3)\right]$.

A quantitative estimate of ZFS splitting can be obtained by switching back to the atomic orbitals,
\begin{align}
\gamma_g&=\langle\Psi_g^{1,2}|\mathcal{H}_{S}|\Psi_g^{1,2}\rangle_{\pm\frac{3}{2}}-\langle\Psi_g^{3,4}| \mathcal{H}_{S}|\Psi_g^{3,4}\rangle_{\pm\frac{1}{2}}\nonumber\\
&=\frac{\gamma_0}{4}\left[\eta_{ad}\langle r_{ad}^{-3}\rangle(1-3\cos^2\theta_{ad})+\eta_{ab}\langle r_{ab}^{-3}\rangle\right],\label{18}
\end{align}
where the $\eta_{ab}=1.443$ and $\eta_{ad}=1.557$ are the respective weight factors of the expectation value $\langle Y_{2,ij}^0/r_{ij}^3\rangle$ originating from total $ad$ and $ab$ pair contributions of MOs after evaluating the determinantal multi-particle wave functions according to Eq.~(\ref{17}) and using the explicit forms of $u$, $v$, and $e_{x,y}$ given in Eq.~(\ref{6}). This equation can also be written in a more familiar form starting from the spin dipole-dipole interaction as
\begin{equation}
\gamma_g=\frac{3}{2}\gamma_0\left\langle \frac{1-3\cos^2\theta}{r_{ij}^3}\right\rangle_{\phi_{ve^2}^{A_2}}\left[S_{z}^2-\frac{1}{3}S(S+1)\right],\label{19}
\end{equation}
where $\gamma_g$ becomes $D[S_z^2-S(S+1)/3]$. So far we assumed all the charge of unpaired electrons is localized on the neighboring carbon atoms. However, as previously reported \cite{Mizuochi2002}, only $62.3\%$ of total charge is localized on the neighboring carbon atoms, and this yields to a reduction of roughly $\tau=(1-(0.377)^2)=0.858$ in $\gamma_g$, i.e. $\gamma_g\rightarrow\tau\,\gamma_g$.

Evaluation of Eq.~(\ref{18}) with these weight factors and structure parameters calculated via DFT, i.e. $r_{ab}=3.3563$\AA, $r_{ad}=3.3567$\AA, and $\theta_{ad}=35.259^\circ$, as well as accounting for the missing charge, results with a ground state ZFS splitting of $2\gamma_g\approx -68$MHz ($D<0$) for an h-site $\textrm{V}_{\textrm{Si}}^-$ defect in good agreement with the experimentally observed values \cite{Kraus_NP14,Carter_preprint15}.

\section{Lambda system}

A $\Lambda$-type three-level system can be created by a magnetic field transverse to the $C_3$-axis. Such a field will mix states in the ground and excited manifolds in different ways. This is because in the ground state manifold there is a small spin-spin splitting between states with $|S_z|=3/2$ and $|S_z|=1/2$, whereas the corresponding states in the excited manifold are split by the much larger spin-orbit interaction. We assume a weak enough magnetic field such that the coupling of the spin states is much smaller than the spin-orbit term $\Delta_e$. This allows the eigenstates in the excited manifold to remain in the form shown in the main text (without a B-field). There are several choices for the composition of the $\Lambda$ system. Below we present some of these options. In all cases, the lower levels are eigenstates of $\hat{S}_x$, which in terms of the states in Table I of the main text are given by
\begin{align}
\Psi_{g,x}^{1} &\simeq \left[\frac{(1-i)}{4}\Psi_{\textrm{g}}^1+\frac{(1+i)}{4}\Psi_{\textrm{g}}^2 + \sqrt{\frac{3}{8}} (\Psi_{\textrm{g}}^3+\Psi_{\textrm{g}}^4) \right]\nonumber
\\
\Psi_{g,x}^{2} &\simeq \left[-\frac{\sqrt{3}}{4}(1+i)\Psi_{\textrm{g}}^1-\frac{\sqrt{3}}{4}(1-i)\Psi_{\textrm{g}}^2- \frac{1}{\sqrt{8}}(\Psi_{\textrm{g}}^3-\Psi_{\textrm{g}}^4) \right]\nonumber
\\
\Psi_{g,x}^{3} &\simeq \left[\frac{\sqrt{3}}{4}(1-i)\Psi_{\textrm{g}}^1+\frac{\sqrt{3}}{4}(1+i)\Psi_{\textrm{g}}^2- \frac{1}{\sqrt{8}}(\Psi_{\textrm{g}}^3+\Psi_{\textrm{g}}^4) \right]\nonumber
\\
\Psi_{g,x}^{4} &\simeq \left[-\frac{(1+i)}{4}\Psi_{\textrm{g}}^1-\frac{(1-i)}{4}\Psi_{\textrm{g}}^2 + \sqrt{\frac{3}{8}} (\Psi_{\textrm{g}}^3-\Psi_{\textrm{g}}^4) \right]\nonumber,
\end{align}
in descending $\langle S_x\rangle$ value, 3/2,1/2,-1/2, -3/2 and where, for simplicity, we have ignored a small correction from the ZFS.

In the first approach for a $\Lambda$ system we can select states with the same weight of $|\uparrow\uparrow\uparrow\rangle$, e.g., states $\{\Psi_{g,x}^{1},\Psi_{g,x}^{4}\}$ or $\{\Psi_{g,x}^{2},\Psi_{g,x}^{3}\}$. In the excited state manifold then we select $\Psi_e^+$, which is defined in the main text and has well-defined projection of spin along the $z$ (or $C_3$) axis, due to the suppression of Zeeman mixing originating from the large SO interaction. The effect is similar to the selection rules in self-assembled quantum dot electron-trion systems under a Voigt B field. Because of this composition of the Lambda system, the polarization of the two transitions is the same. The frequency however is different, and that degree of freedom can be used as the `handle' with which to the emitted photon can be manipulated.

An alternate scheme for a $\Lambda$ system is to select as lower levels the eigenstates of $\hat{S}_x$ with eigenvalues -1/2 and 3/2, $\Psi_{g,x}^{3}$ and $\Psi_{g,x}^{1}$. In the excited state manifold the relevant state is then $(\Psi_{\textrm{q2}}^7+\Psi_{\textrm{q2}}^8)/\sqrt{2}$, i.e., again mixing to the states with different spin projection along $z$  has been ignored in the excited manifold due to the large SO splitting. We note that here the two transitions have the same polarization but, unlike the scheme above, different dipole moments, originating from the different coefficients of $\Psi_{\textrm{g}}^3$, $\Psi_{\textrm{g}}^4$ in the states $\Psi_{g,x}^{3}$ and $\Psi_{g,x}^{1}$.
